\documentclass[prb,superscriptaddress,aps]{revtex4}
\usepackage[english]{babel}
\usepackage{amsmath}
\usepackage{graphicx}
\usepackage{epsfig}
\usepackage{subfig}
\usepackage{dcolumn}
\usepackage{bm}

\begin{document}

\preprint{}
\title{On the transition between the Weibel and the whistler instabilities}
\author{L. Palodhi, F. Califano, F. Pegoraro}
\affiliation{Phys. Dept. Pisa University and CNISM, Pisa, Italy }

\begin{abstract}
The transition between non resonant (Weibel-type) and resonant (whistler) instabilities is investigated numerically in   plasma configurations  with an ambient magnetic field  of increasing amplitudes.  The Vlasov-Maxwell system is solved in a  configuration where the fields have three components but depend only on one coordinate and on time.   The nonlinear evolution of these instabilities  is shown to lead to the excitation of electromagnetic and electrostatic modes at the first few harmonics of the plasma frequency and, in the case of a large ambient magnetic field, to a long-wavelength, spatial modulation of the  amplitude of the magnetic field generated by the whistler instability.

\end{abstract}

\maketitle

\section{Introduction}

Electron  distribution functions that are anisotropic in  phase space  are a common feature of collisionless plasmas  both  in space and in the laboratory and the investigation of the processes through which these distributions relax  is of primary interest.   In fact,   the free energy  that is made available by the unbalance  of the particle ``temperatures''  in the different directions  can be transferred, depending on the plasma conditions,  to quasistatic magnetic fields,  to electromagnetic  or electrostatic coherent structures or to particle acceleration.

The anisotropy of the electron distribution function in an unmagnetized plasma can give rise to the onset of the well known  Weibel instability\cite{Weibel} which  generates  a quasistatic magnetic field.

If  a magnetic field  is already present  in the  plasma, the Weibel instability driven by the {anisotropy of the electron energy distribution} turns into  the so called  whistler instability\cite{WIn} in which case  circularly polarized whistler  waves are generated by the relaxation of the electron distribution function.  Whistler waves  are  actually ubiquitous in plasmas and their generation has been extensively studied in recent years  in the laboratory  (see e.g. Ref.\onlinecite{stenzel}).  Whistler instabilities have been reported in space\cite{Wsp} where bursts of whistler mode magnetic noise are found to be present in the magnetosphere, close to the magnetopause  and are also a likely source of several different magnetospheric fluctuations including plasmaspheric hiss and magnetospheric chorus. These waves propagate along the ambient magnetic field in the frequency range of $\omega_{ci} \le \omega_{r} \le \omega_{ce}$, where $\omega_{ce}$ and $\omega_{ci}$ are the electron and proton cyclotron frequencies.
A sufficiently large  temperature anisotropy in a magnetized plasma is the most commonly observed mechanism for the generation of whistler waves, but additional mechanisms, such as trapped particle loss cone distributions,   can also generate these modes.

The linear dispersion relations of these instabilities and how the Weibel instability merges into the whistler instability  have been studied in the literature\cite{WW}. It has been shown\cite{WW}  that  ``Weibel-type'' whistler modes   occur  in plasma conditions where the following independent inequalities are satisfied
\begin{equation}
\omega_r \leq \gamma  \, ,     \qquad     ( {v_{the,\perp}}/{c})^2 > ({\omega_{ce}}/{\omega_{pe}})^2
\label{cond1}  ,
\end{equation}
where $\omega_r $ and $\gamma$ are the real and imaginary parts on the mode frequency respectively, and  depend on the parallel (${v_{the,||}}$)
 and on  the perpendicular (${v_{the,\perp}}$) thermal velocities of the electrons, and $\omega_{pe}$ is the electron Langmuir frequency.
 On the contrary, if the  above threshold,  which we rewrite as
 \begin{equation}
\omega_{ce} > \omega_{pe}\sqrt{(A+1)}{v_{the,||}}/{c},
\label{cond2}
\end{equation}
 is passed,
 the growth rates  of the whistler instability depart considerably from those of the Weibel instability. In this case the mode  frequency $\omega_r $ becomes comparable or greater than its growth rate $\gamma$. Here  the anisotropy parameter $A$  is defined by  $A   \equiv   (v_{the,\perp}/v_{the,||})^2 -1$.

In the present  paper  we consider the long term collisionless relaxation process of an anisotropic electron distribution focusing our attention on the nonlinear features of the transitional regime  between  the Weibel and the whistler instabilities
and on the onset of secondary modes driven by the perturbations in the electron distribution function arising from the nonlinear development  of the primary  Weibel and  whistler instabilities.

The analysis  presented here  is performed by solving numerically the collisionless Vlasov equation for electrons and  protons coupled to Maxwell's equations, in a restricted geometrical configuration   (1D-3V) where all vector quantities are three dimensional  but depend on one coordinate only, which is chosen to be the $x$ coordinate, and on time $t$.  More specifically, we consider a homogeneous initial plasma configuration with isotropic protons and  a bi-Maxwellian electron distribution  with $T_{\perp}>T_{||}$  (where ``parallel''  refers to the $x$ direction) in the presence of a uniform  ambient magnetic field $B_x$.  The ratio between the proton temperature $T_i$ and the parallel electron temperature  $T_{||}$ is taken equal to one. Vector quantities must be taken to be three dimensional because whistler waves are circularly polarised in the $y$-$z$ plane and in addition we need to include longitudinal electric fields and velocities along $x$.   This contrasts the treatment given for the Weibel instability (with ambient magnetic field $B_x =0$) given in Ref.\onlinecite{lopa1} where a 1D-2V configuration was considered.

 At low ambient magnetic field, the nonlinear development of the whistler instability is similar to  that of  the Weibel instability, since the perturbed magnetic fluctuations are much larger  than the ambient magnetic field.   However, when the ambient magnetic field is sufficiently strong, the  long term nonlinear   behaviour  of the instability  changes drastically.
  In both limits the isotropization of the electron distribution function  due to the onset of the instabilities  is accompanied by the development of high frequency (Langmuir wave) electron density modulations\cite{lopa1}.   These density modulations are forced  by the spatial modulation of the magnetic energy density  at   wave numbers   roughly two times  the wave numbers  of the most unstable Fourier components of  the primary instabilities.  However these density modulations become weaker as the ambient magnetic field is increased. In addition a purely kinetic effect is present in the small $B_0$ limit  where the development of the Weibel  instability  leads to  strong deformations  of the electron distribution function in phase space. These deformations  have been shown in a 1D-2V
configuration\cite{lopa1}  to   arise from the  differential rotation in velocity space of the electrons
around  the magnetic field $B_z$ produced by the Weibel instability  and to generate  short wavelength Langmuir modes that form highly localized electrostatic structures corresponding to jumps of the electrostatic potential. These kinetic effects  become weaker and eventually disappear as the ambient magnetic field is increased.
\\In the case of the whistler  instability,   low-frequency density modulations involving the proton dynamics  occur and  generate {\it soliton} type structures known as  {\it whistler oscillitons}\cite{Sauer}. Oscillitons  are coherent nonlinear structures that occur in dispersive  media for which the curves of the phase and  group velocities cross for a finite value of the oscillation wavenumber $k$, see Ref.\onlinecite{Sydora}.   Whistler oscillitons  are of special importance as they have been  invoked in order to describe coherent wave emission in the whistler frequency range  observed in the Earth's plasma environment\cite{Sauer,Sydora}.

\section{Numerical set up}

{In the numerical solutions  of the Vlasov-Maxwell system presented below,  all the parameters are normalized as follows. We use the plasma frequency $\omega_{pe}$ and the velocity of light $c$ as characteristic frequency and velocity; therefore the electron skin depth $c / \omega_{pe}$ is the characteristic length scale. The electric and magnetic field are normalized to ${\bar E} = {\bar B} = m_e c \omega_{pe} / e$ and, finally, the electron temperature is normalized to $m_e c^2$. }

The {initial} electron anisotropy is kept fixed, $T_y = T_z = T_{\perp}$, \,  $ T_{||} = T_{x} $, \, $ T_{\perp}/T_{||}= 12$,  while the value of $B_x$ is  varied  such that the corresponding electron  cyclotron frequency, normalized on the electron Langmuir frequency  and indicated in the following simply by $B_0$,  changes from
 $B_0 =0$ to $B_0 =0.5$.   In these numerical integrations Ref.\onlinecite{Mangeney} the ratio $\beta_{||}\equiv  v_{the,||}^2/B_0^2$  plays the role of  a control parameter. Here the parallel thermal velocity $v_{the,||}$ is normalized to $c$ and is chosen to be equal to $0.02$.  The following values of $B_0$ are chosen, $B_0 = 0,\, 0.02,\, 0.1,\,$ and $ 0.5 $, corresponding to $\beta_{||} = \infty,\, 1,\, 4\times 10^{-2},\, $ and $1.6 \times 10^{-3}$ respectively.

The number of grid points in  coordinate space is $N_x = 1024$  and the length of the system is $L_x = 10\pi$ in units of the electron skin depth $d_e \equiv c/\omega_{pe}$. The total number of grid points in
velocity space is $N_{v_{x}} = 101, N_{v_{y }}= N_{v_{z}} = 71$. The system is evolved up to the time $t = 1200\, \omega_{pe}^{-1}$ with time step $10^{-3} \leq dt \leq 10^{-2}$.   The total number of perturbed modes   at $t=0$ is $N = 220$.
The spatial grid spacing is $dx = 0.03$. Henceforth $k_{max}\lambda_D \ll 1$  and $k_{max} d_e \sim 1$, where $k_{max}$  is the   the most unstable wave number  and $ \lambda_D$ is the electron Debye length computed with the  temperature $T_{||}$.
The numerical integrations are initiated  in a homogeneous plasma configuration with isotropic protons  and Maxwellian electrons with different parallel and perpendicular temperatures  by imposing a low amplitude magnetic field noise, randomly oriented in the $y$-$z$  plane at different $x$ positions:
\begin{equation}
B_{y} (x,t=0)= \sum^{N}_{n=1} a_{in} \ [\cos(k_{n}x+\psi_n)], \qquad B_{z} (x,t=0)= \sum^{N}_{n=1} b_{in} \ [\cos(k_{n}x+\phi_n)],
\label{IC2}
\end{equation}
with $a_{in}$ and $ b _{in}$  the initial "small" ($\sim 10^{-4}$) amplitudes,  $\psi_n$ and $\phi_n$ random phases, {$n$ the wave number and $\mathbf k_{n}=2\pi n/L_{x}$ the corresponding wave vector.}

In Table \ref{table1} we list the main parameters of the numerical runs, together with the analytical (kinetic)  value,
$\gamma_{max}^{Ana}$ of the  growth rate of the most unstable mode, $k = k_{max}$, as obtained, { after some simplifications}, from Ref.\onlinecite{WW}, the corresponding value,  $\gamma_{max}^{Num}$, as obtained by the numerical integration  of the Vlasov-Maxwell system and the value of the ratio  $ {\hat B}_z $ between   the amplitudes of  the $z$ component of the magnetic field produced by the instability at saturation and that of the ambient magnetic field along $x$.
\begin{table}[!h]
\centering
\begin{tabular}{c c c c c c c c c}
\hline
\\
Instability & ${\bf{B_{o}}}$  &     $\omega_{pe}/\omega_{ce}$  &  $k_{max}$       &   $\gamma_{max}^{Ana}$  &   $\gamma_{max}^{Num}$     &  $ {\hat B}_z $  &  $\beta_\|$  & $\beta_{\perp}$\\ [0.5ex]
\hline
\\
Weibel & 0.0             &             -               &     1.0         &     0.034           &     0.037               &      -             &    -        &   -              \\
Whistler & 0.02            &            50               &     1.2         &     0.037           &     0.038               &     2.25           &    1        &   11.75        \\
& 0.10            &            10               &     1.2         &     0.031           &     0.035               &     0.56           &  0.04       &   0.50           \\
& 0.50            &             2               &     2.0         &     0.025           &     0.020               &     0.06           &  0.0016     &   0.02            \\
\\
\hline
\end{tabular}
\caption{\small  Main parameters of the numerical runs, the corresponding  numerical and analytical growth rates for the most unstable mode corresponding to $k = k_{max}$ and the normalized amplitude at saturation  of  the $z$ component of the magnetic field produced by the instability.}
\label{table1}
\end{table}
\,
In Table \ref{table2} the real part of the frequency of the whistler  unstable mode  is shown  for different values of $k$ in the ``large'' ambient magnetic field case $B_0= 0.5$.  For each $k$ three values of  the real part of the frequency    are shown. Two are analytical:   $\omega_r^{Ana}(kinetic)$ is obtained, after some simplifications,  from   Ref.\onlinecite{WW} where a kinetic approach  is used, and   $\omega_r^{Ana}(fluid)$  is obtained  from  Ref.\onlinecite{VAL} where a fluid approach  is used,  while $\omega_r^{Num}$ is obtained by    numerical integration of the Vlasov-Maxwell system.\,
The comparison between the analytical and the  numerical  frequencies and growth rates shows  good agreement.  The numerical values of the frequencies and growth rates shown  in Tables  \ref{table1}-\ref{table2} are obtained  by considering only the linear part of the numerical integration. It should be noted that the comparison for the real frequencies is shown only for ${\bf B}_0 = 0.5$. For the other values of $B_0$ in Table \ref{table1},  the  oscillation period   would turn out to be  longer than the total duration  time of the linear regime and  it is  thus  difficult  to estimate numerically. \\

\begin{table}[!h]
\centering
\begin{tabular}{c c c c}
\hline\\
$k$                 &      $\omega_r^{Ana}(kinetic)$     & $\omega_r^{Num}$  &       $\omega_r^{Ana}(fluid)$      \\ [0.5ex]
\hline
\\
0.2                 &        0.02                   &       0.02        &               0.02                \\
0.4                 &        0.08                   &       0.08        &               0.08                 \\
0.6                 &        0.18                   &       0.13        &               0.13                 \\
0.8                 &        0.32                   &       0.22        &               0.20                 \\
1.0                 &        0.37                   &       0.25        &               0.25                  \\
1.6                 &        0.43                   &       0.40        &               0.36                  \\
2.0                 &        0.45                   &       0.42        &               0.40                   \\
2.4                 &        0.46                   &       0.42        &               0.42                   \\
3.0                 &        0.47                   &       0.45        &               0.45                   \\
\\
\hline
\end{tabular}
\caption{\small  Mode frequency as a function of $k$  for  $B_0 = 0.5$ - Numerical and analytical values.}
\label{table2}
\end{table}
In Table \ref{table3} the real part of the   frequency of the   most unstable  whistler  mode, as derived analytically in Ref.\onlinecite{VAL} is shown for different  values of the ambient magnetic field.
\begin{table}[!h]
\centering
\begin{tabular}{c c c c}
\hline\\
Instability                  &      $B_0$               &      $k_{max}$                 & $\omega_r^{Ana.}$      \\ [0.5ex]
\hline
\\
whistler                     &       0.02               &       1.20                     &        0.02            \\
                             &       0.10                &      1.20                     &        0.08             \\
                             &       0.50               &       2.00                     &        0.47             \\

\\
\hline
\end{tabular}\caption{\small  Frequency of the most unstable  whistler mode for different  ambient magnetic fields }
\label{table3}
\end{table}

\section{Whistler modes: transition from weak to strong ambient magnetic field }

From  Table~\ref{table1} we see that the growth rate of the whistler instability decreases  as
the amplitude of the ambient  magnetic field increases, while the mode number $k_{max}$,
at which the maximum growth rate of the whistler instability is obtained,  increases.

By referring to the inequalities (\ref{cond1})
 we  see that for  small ambient magnetic field amplitudes  both  inequalities   are satisfied so that  the mode frequency and its  growth rate  are approximately those obtained  in Ref.\onlinecite{lopa1} for the Weibel  instability (zero ambient magnetic field).  On the contrary,  as  the  ambient magnetic field amplitude is increased, the system goes through the threshold condition (\ref{cond2}) and at the same time  the frequency of the whistler waves becomes comparable to, or larger than, their growth rate ($\omega_r > \gamma_{max}$).
Hence these numerical  runs describe  the  transition regime between Weibel-like,  low ambient magnetic field,  instabilities  and strong ambient magnetic field whistler instabilities.

 \subsection{Electromagnetic fields}

\begin{figure}[!h]
\vspace{0.cm}
\centerline{\hspace{-0.2cm}\psfig{figure=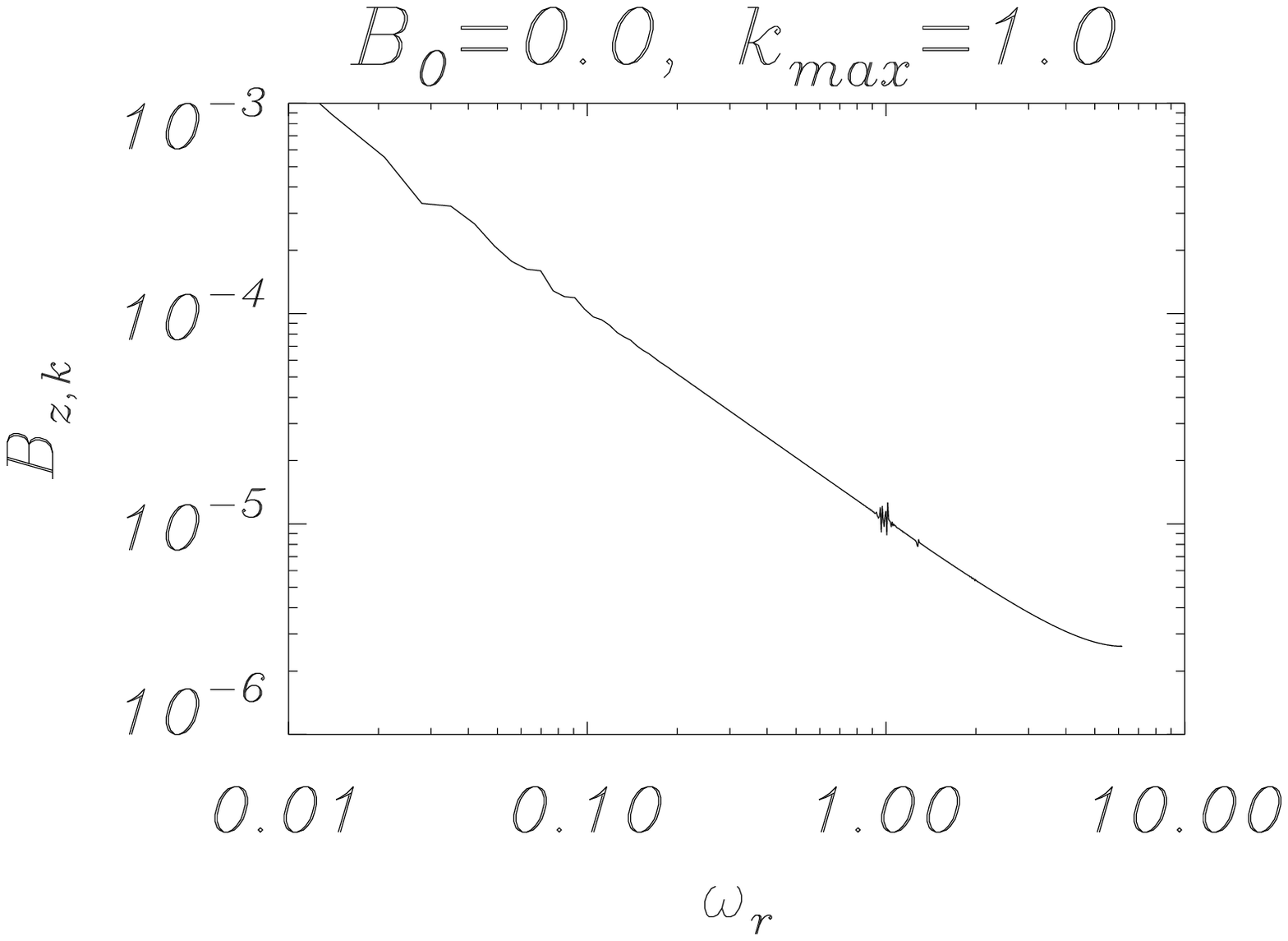,height=4.5cm,width=6cm}
\hspace{-0.6cm}\psfig{figure=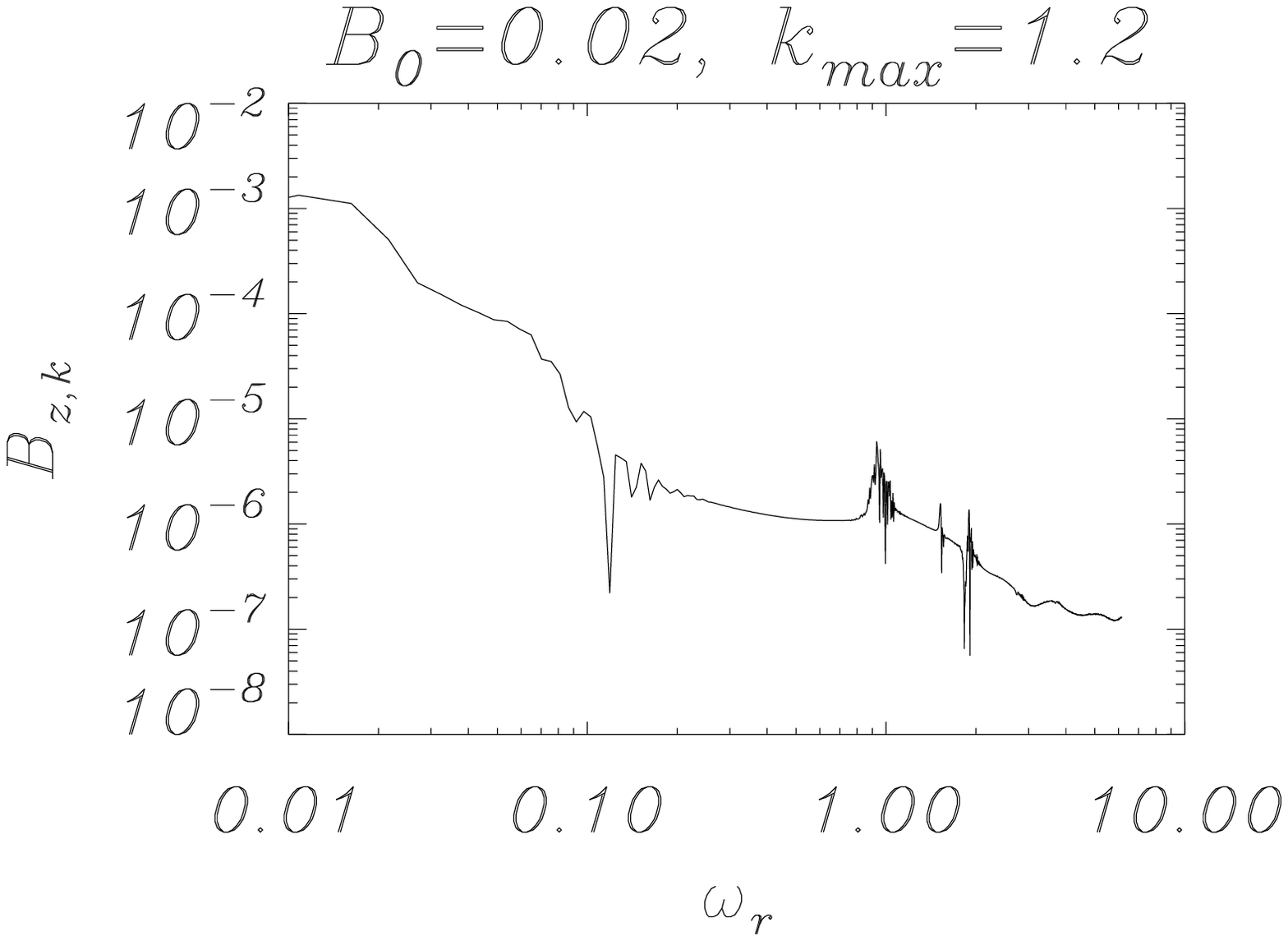,height=4.5cm,width=6cm}}
\vspace{-0.1cm}\centerline{\hspace{-0.2cm}\psfig{figure=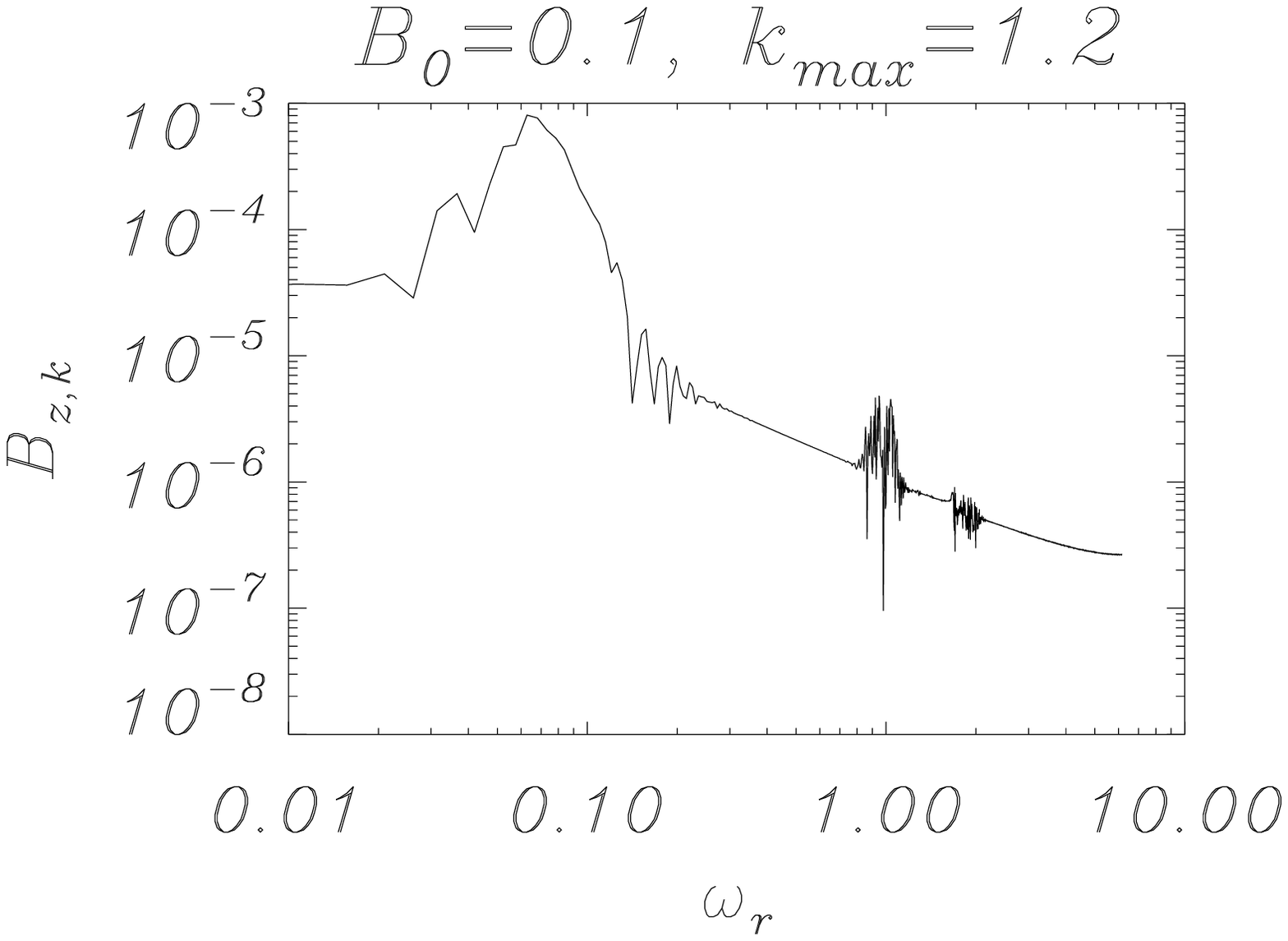,height=4.5cm,width=6cm}
\hspace{-0.6cm}\psfig{figure=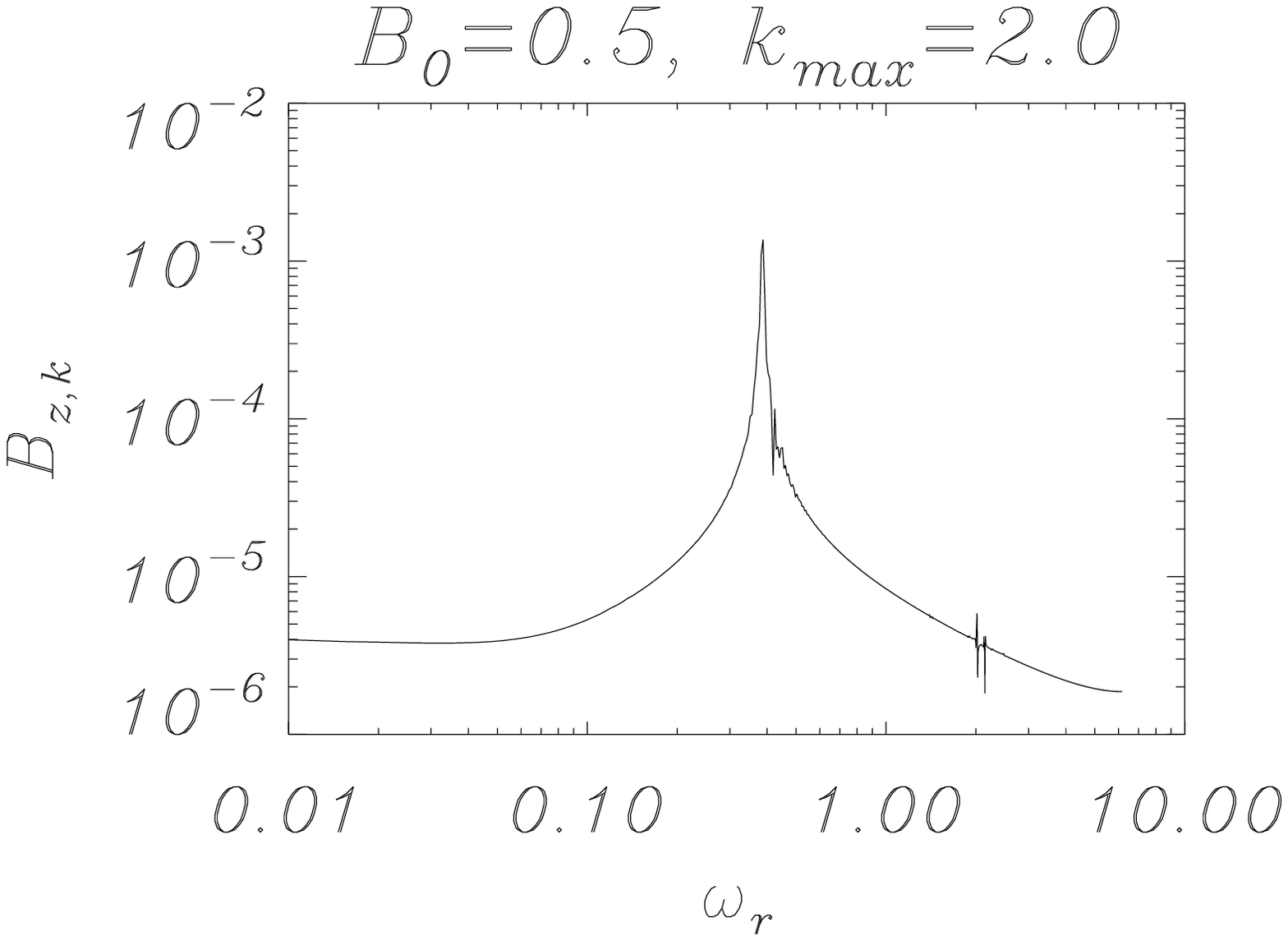,height=4.5cm,width=6cm}}
\vspace{-0.4cm}
\caption{\small Frequency spectrum of the magnetic field  component $B_{z,k}$ for $B_0 = 0, 0.02, 0.1, 0.5$ at the corresponding  most unstable mode $k =k_{max} = 1.0,1.2, 1.2, 2.0$. Note the change in the amplitude scale in the different frames.}
\label{Fig1}
\end{figure}

The frequency spectrum  of the $z$ component of the perturbed magnetic field  is shown in Fig.\ref{Fig1}  for ${B_0=0,\, 0.02\, ,0.1,\, 0.5}$.  For each value of  the ambient magnetic field $B_0$    the most unstable spatial Fourier mode (corresponding to $k = k_{max}$ in Table \ref{table1}) is chosen.
The frequency spectrum is obtained  by integrating  over time over the whole simulation   time ($t = 1200\, \omega_{pe}^{-1}$).
\noindent
We see  a  strong peak at $\omega_r = 0.45$ in the spectrum of the magnetic field with $B_0 = 0.5$
(bottom right frame) which corresponds to the frequency of the whistler wave for  $k = k_{max} =2$.
This wave is observed at frequencies $\omega_r=0.015$ and $\omega_r = 0.07$ for  $B_0=0.02$ and for
$B_0 = 0.1$, respectively. The whistler  peak gets sharper, the stronger the ambient magnetic field.
This reduction of the mode "line width"  is consistent with the decrease of the mode growth rate at larger ambient magnetic field values. For comparison, the spectrum  for the case  $B_0=0$ (Weibel case)  is also shown:  no peak  is observed as  expected since in an unmagnetized plasma the instability is purely growing.
\begin{figure}[!h]
\vspace{0.cm}
\centerline{\hspace{-0.2cm}\psfig{figure=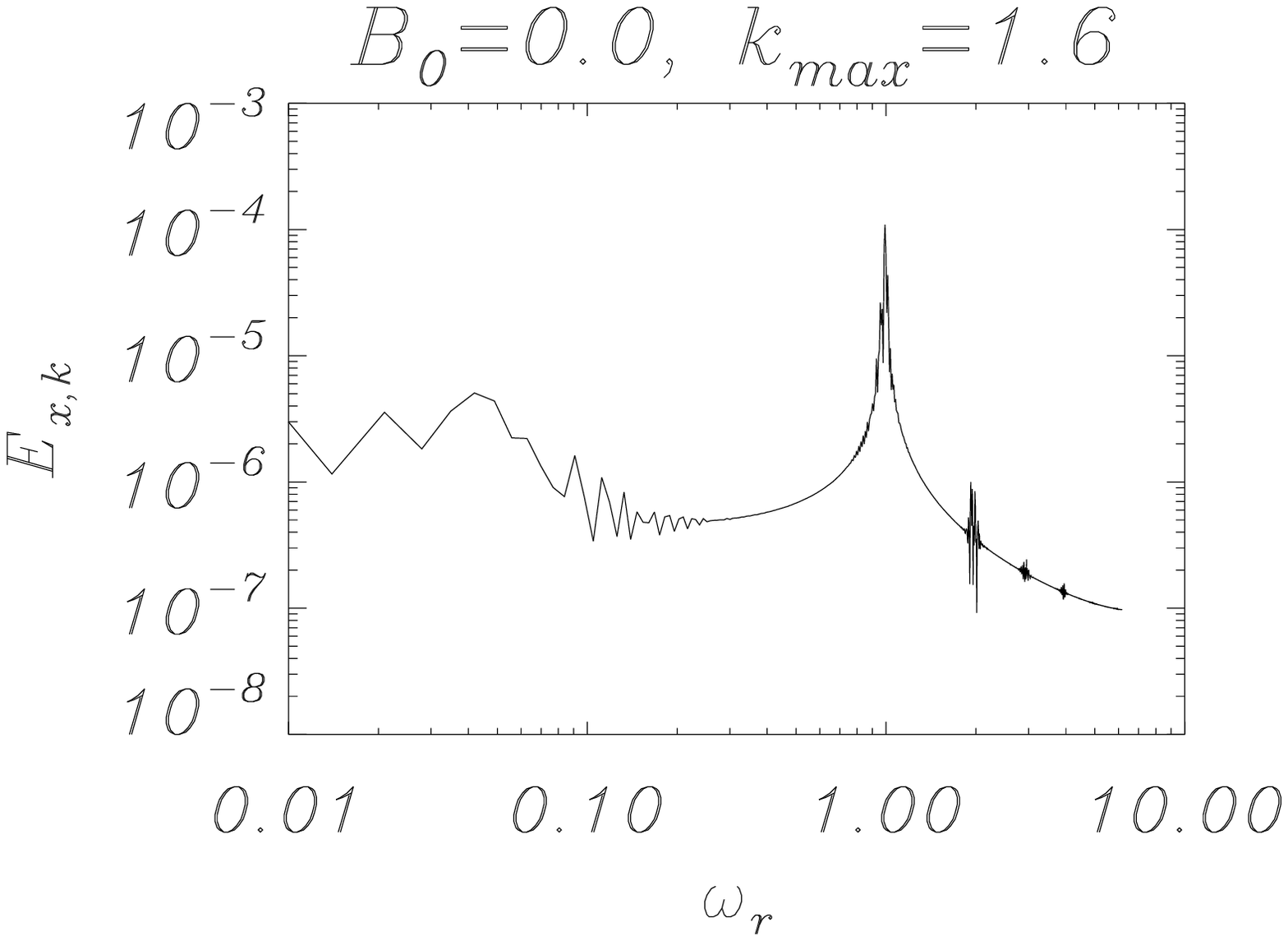,height=4.5cm,width=6cm}
\hspace{-0.6cm}\psfig{figure=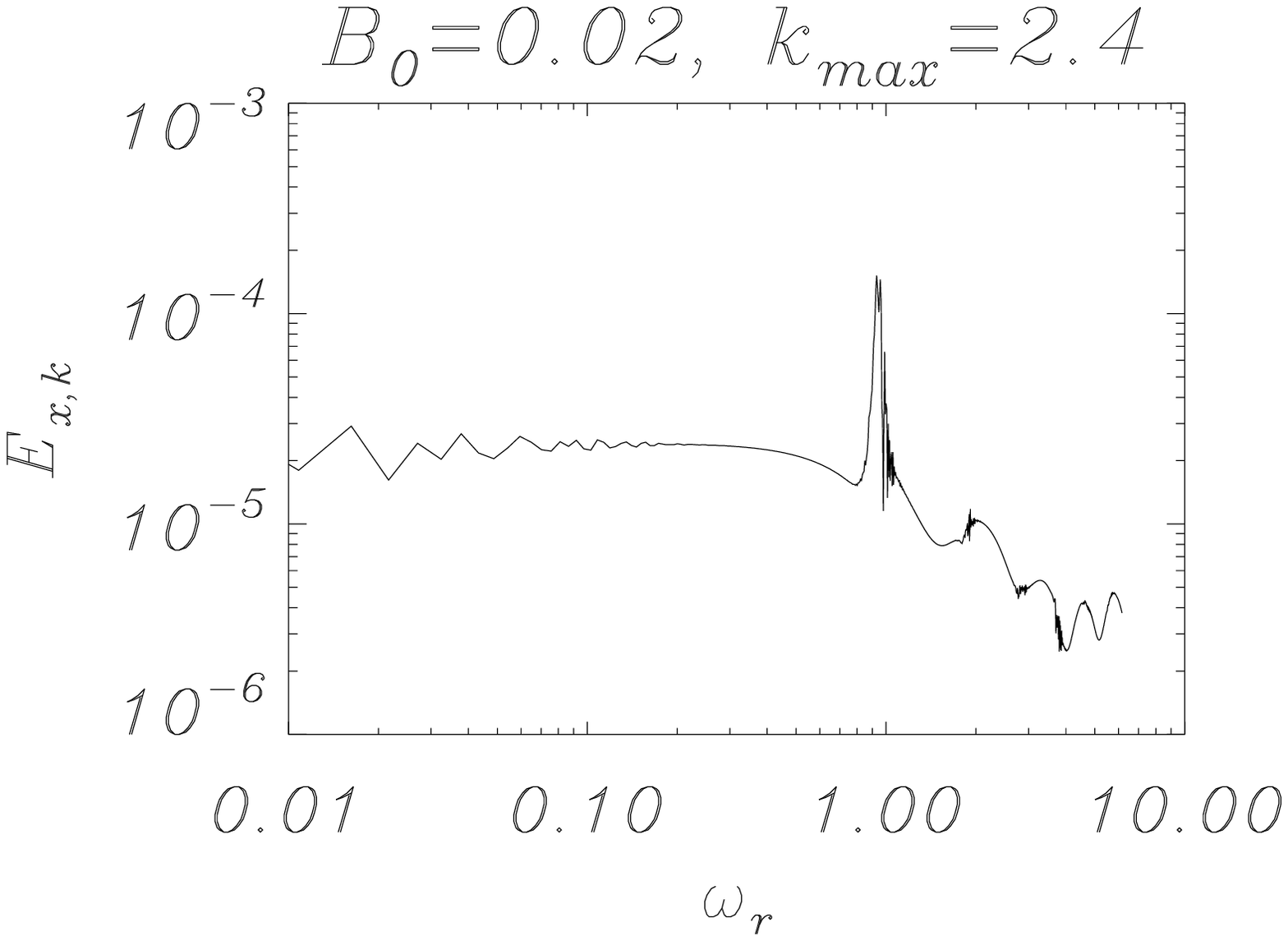,height=4.5cm,width=6cm}}
\vspace{-0.1cm}\centerline{\hspace{-0.2cm}\psfig{figure=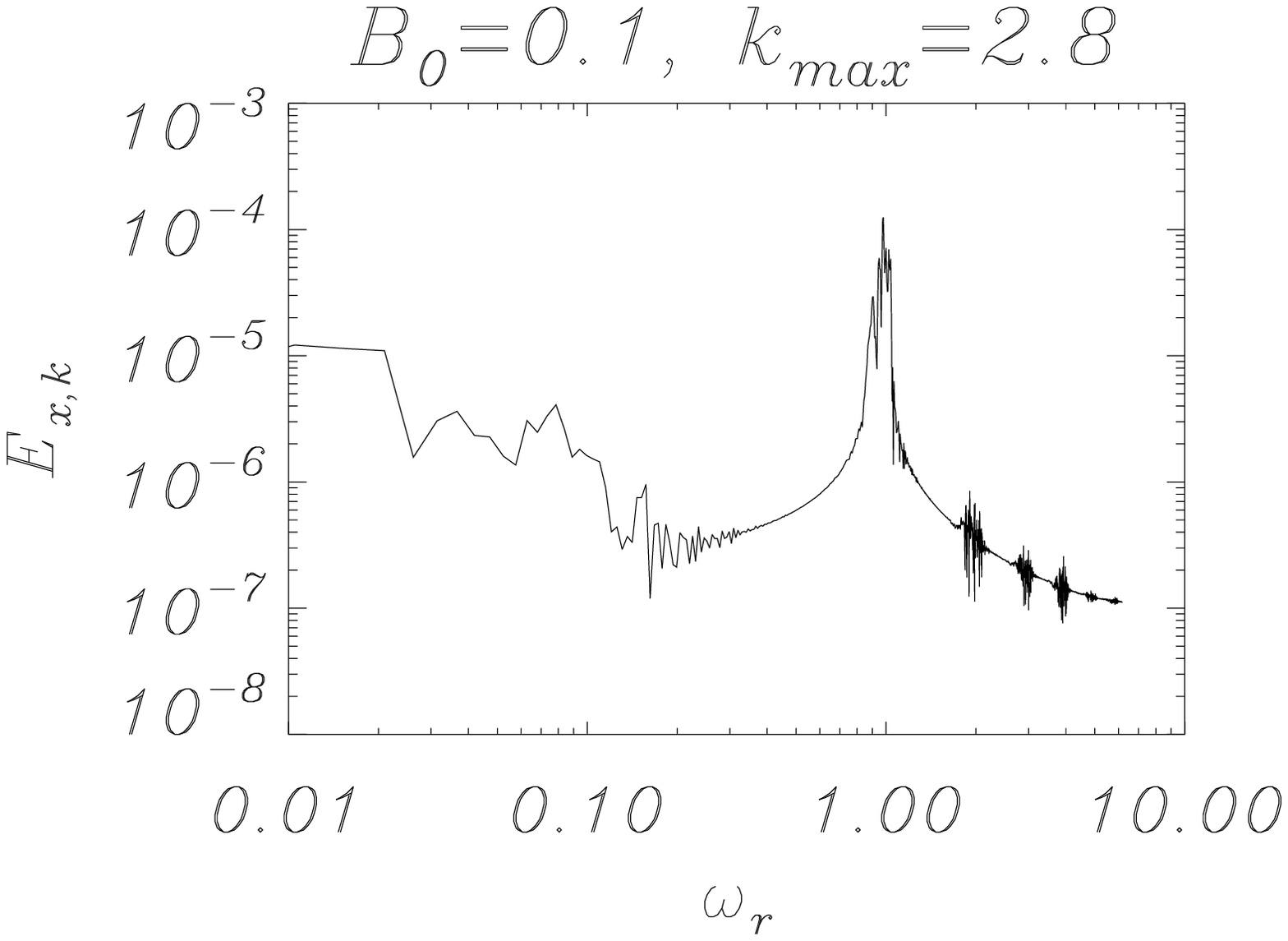,height=4.5cm,width=6cm}
\hspace{-0.6cm}\psfig{figure=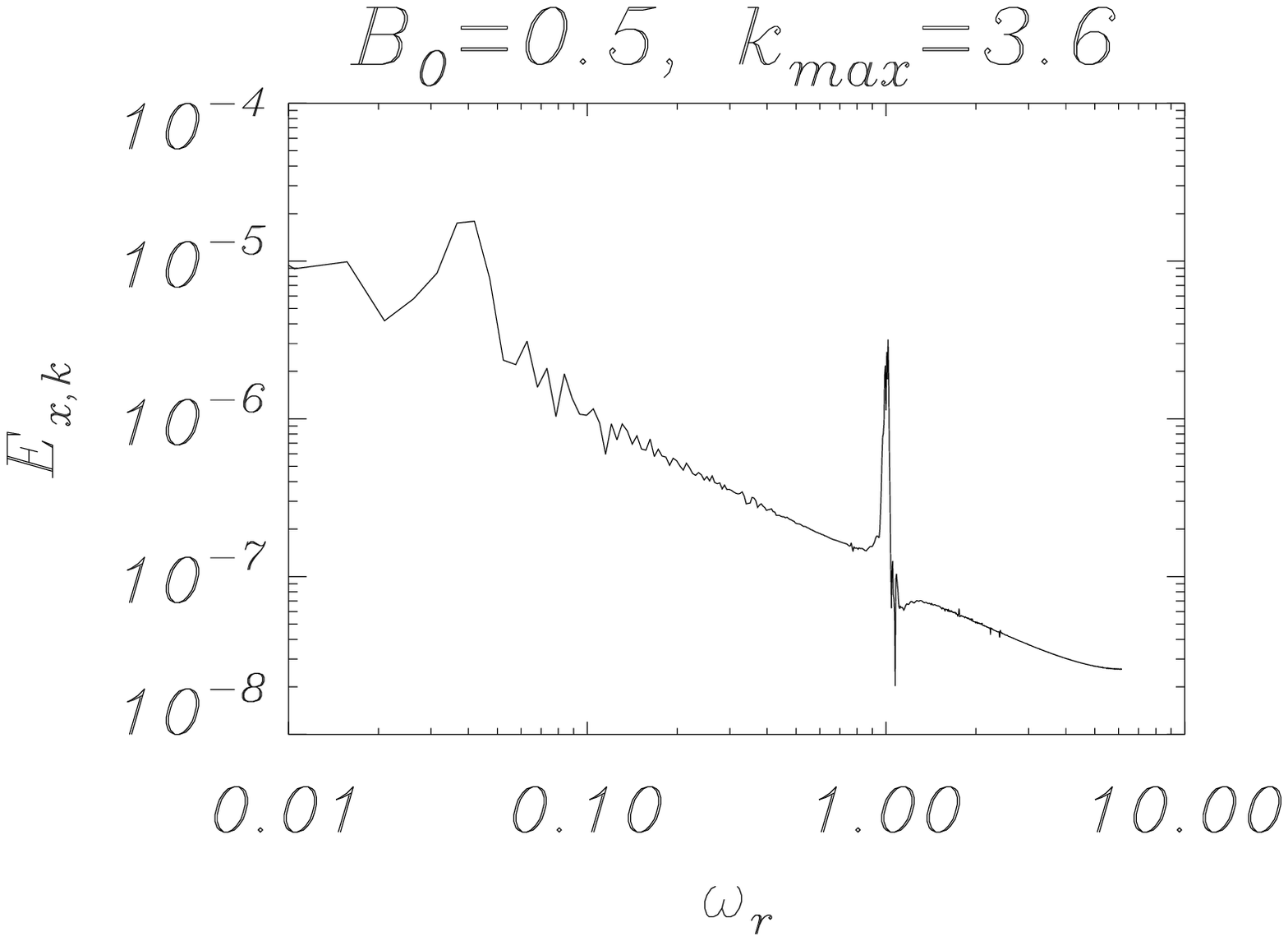,height=4.5cm,width=6cm}}
\vspace{-0.4cm}
\caption{\small Frequency spectrum of the electrostatic field  component $E_{x,k}$ for $B_0 = 0, 0.02, 0.1, 0.5$ at the corresponding  most unstable mode $k =k_{max} = 1.6,2.4, 2.8, 3.6$. Note the change in the amplitude scale in the different frames. }
\label{Fig2}
\end{figure}
 \noindent
We observe that both  in the Weibel and in the  whistler cases   the  frequency spectrum contains waves that  propagate with a frequency equal to the plasma frequency, $\omega_{pe}=1$ in the dimensionless units used. In the two intermediate cases, $B_0=0.02$ and $B_0=0.1$ we also see  waves at $2\omega_{pe}$.
\begin{figure}[!h]
\centerline{\hspace{-0.2cm}\psfig{figure=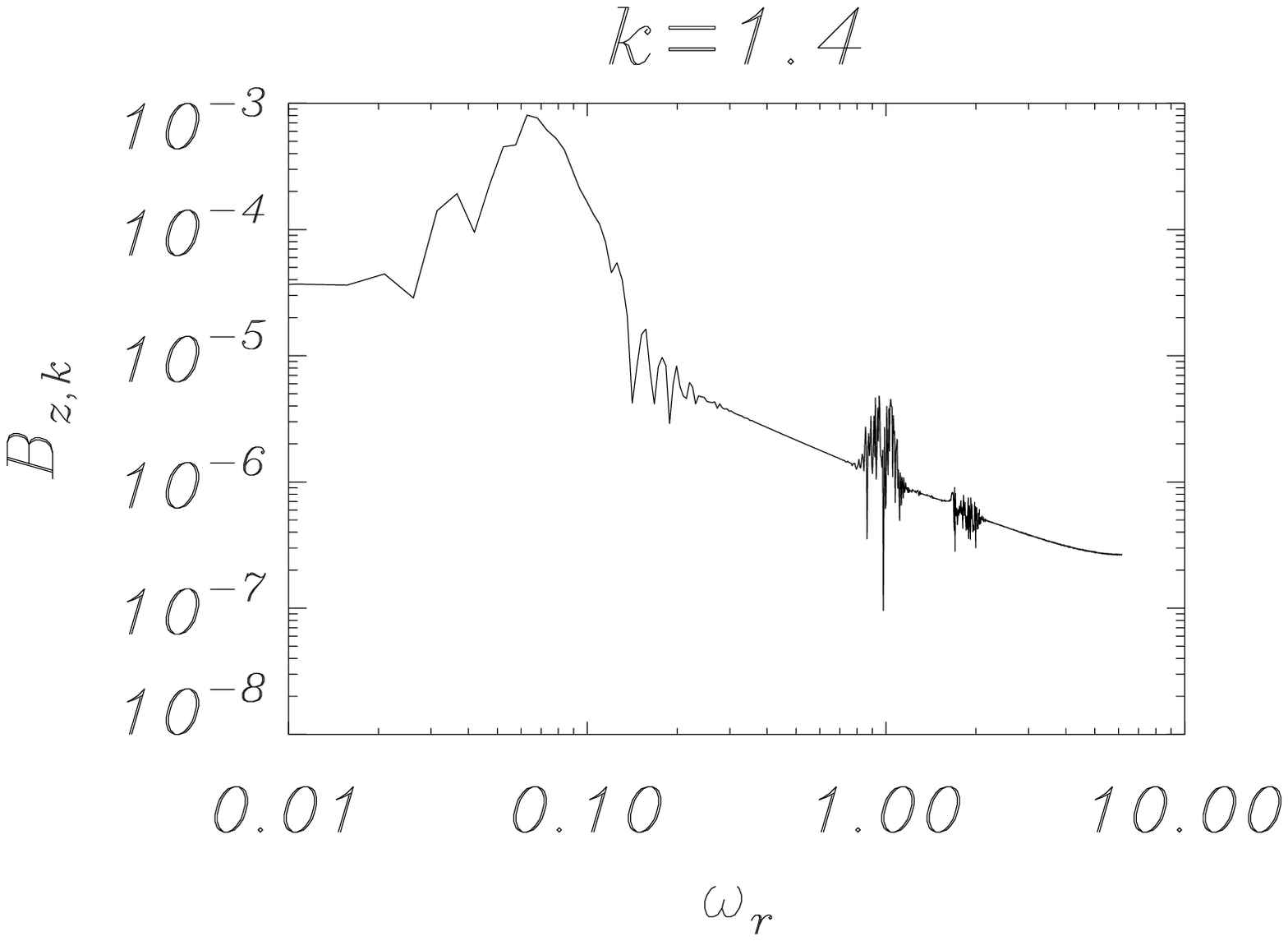,height=4.6cm,width=5.7cm}
\hspace{-0.4cm}\psfig{figure=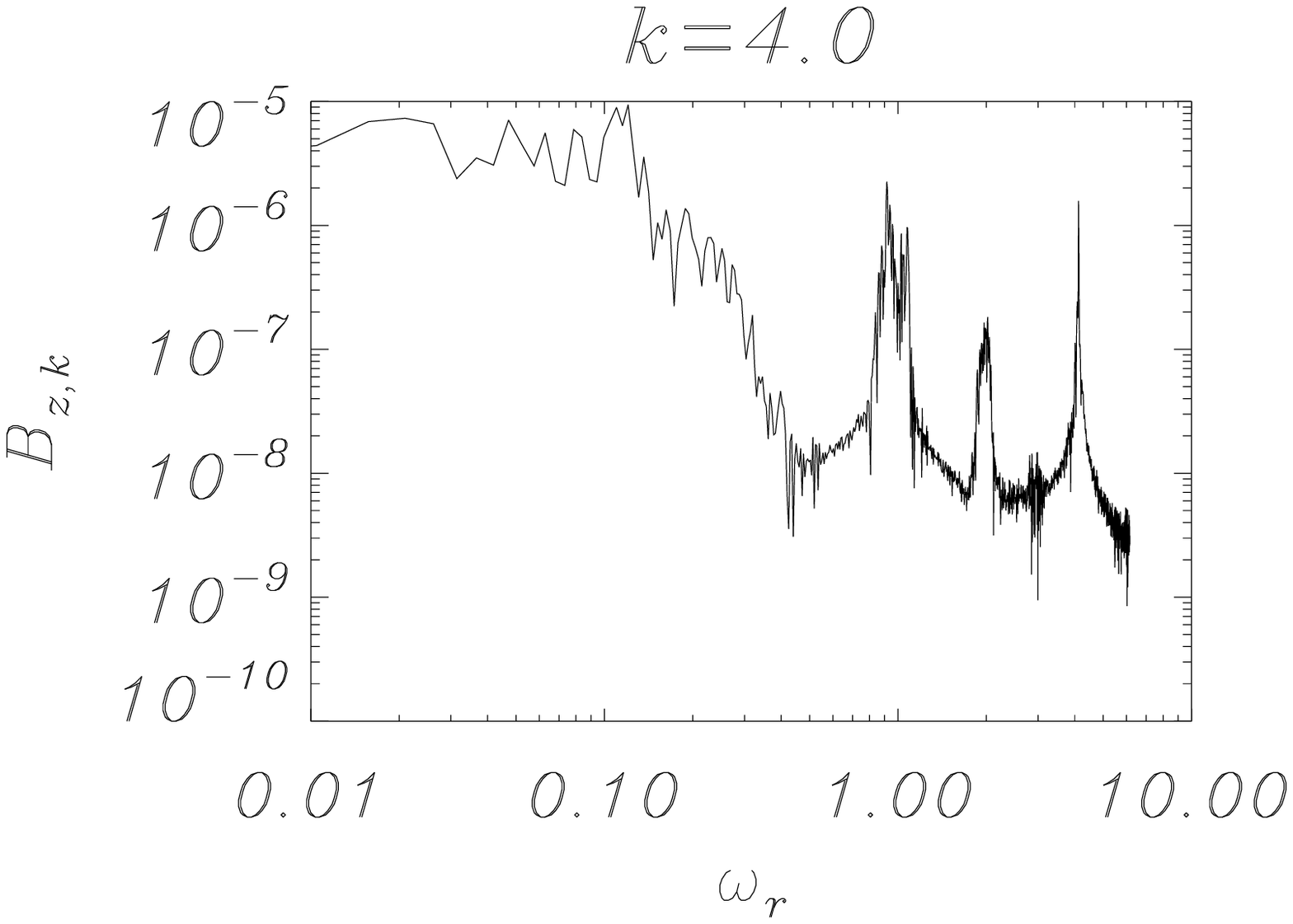,height=4.6cm,width=5.7cm}
\hspace{-0.4cm}\psfig{figure=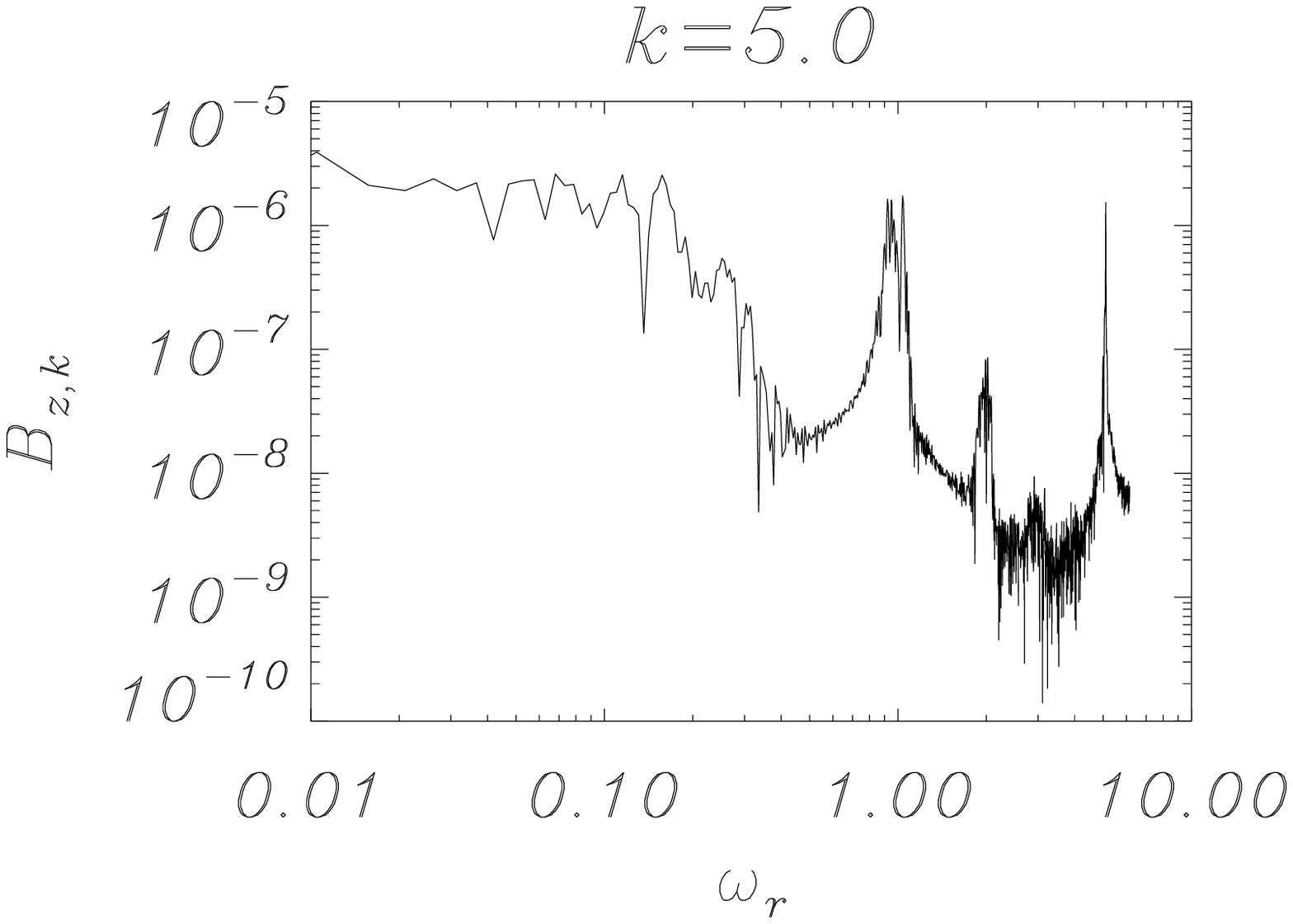,height=4.6cm,width=5.7cm}}
\vspace{-0.4cm}
\caption{\small Frequency spectrum of the Fourier component of the magnetic field, $B_{z,k}$ for $B_0 = 0.1$ at  $k = 1.4,4.0,5.0 $. Note the change in amplitude scale.}
\label{Fig3}
\end{figure}
\noindent
Wave emission with characteristic frequency  close to twice the electron plasma frequency are observed near the Earth's bow shock\cite{Gurnett1975} and also from type II and III solar radio bursts\cite{Goldman}.  Different mechanisms for the generation of such kind of  wave emissions have been  discussed in the literature, see Ref.\onlinecite{Malaspina} and references therein.   Based on the  frequency  spectrum of the longitudinal electric field $E_x$ shown in Fig.\ref{Fig2},  a plausible    mechanism for  the $2\omega_{pe}$ emission in our system  can be  the  coupling between  forward and backward propagating Langmuir waves that  generate  an electrostatic mode at $2\omega_{pe}$ which is then converted into an electromagnetic wave by a nonlinear mode coupling process.  The amplitude  of the $2 \omega_{pe}$ component of the perturbed magnetic field  is  much smaller than that of the waves radiated at the plasma frequency.   This small amplitude electromagnetic emission at frequencies close to the harmonics of the plasma frequency is not limited to the modes at $k=k_{max}$ shown in Fig.\ref{Fig1} and extends over a range of  different values of $k$.  Within this range   emission  at $3\omega_{pe}$ is also observed (Fig.\ref{Fig3})  at short wavelengths.

\begin{figure}[!h]
\centerline{\hspace{-0.2cm}\psfig{figure=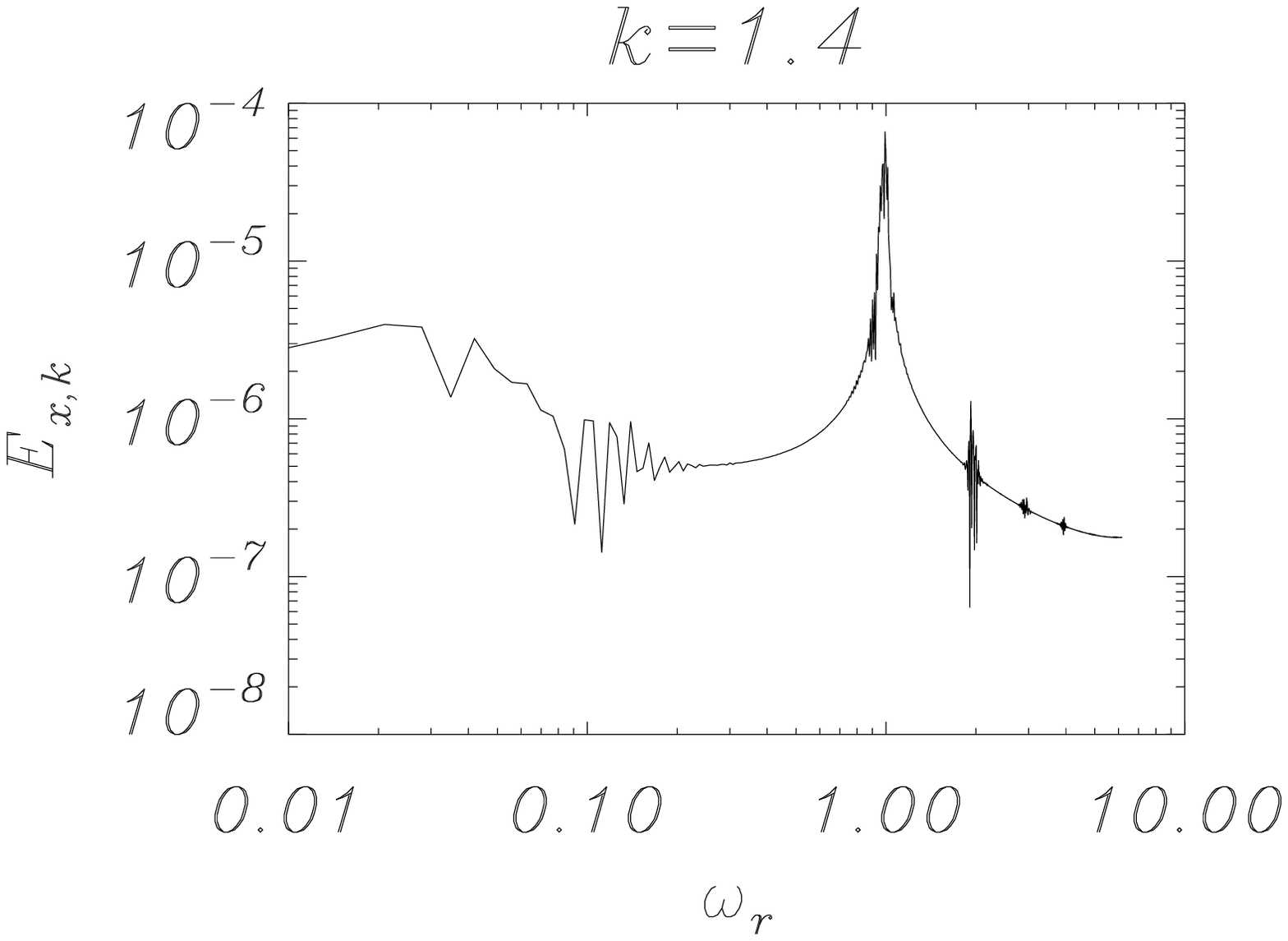,height=4.6cm,width=5.7cm}
\hspace{-0.4cm}\psfig{figure=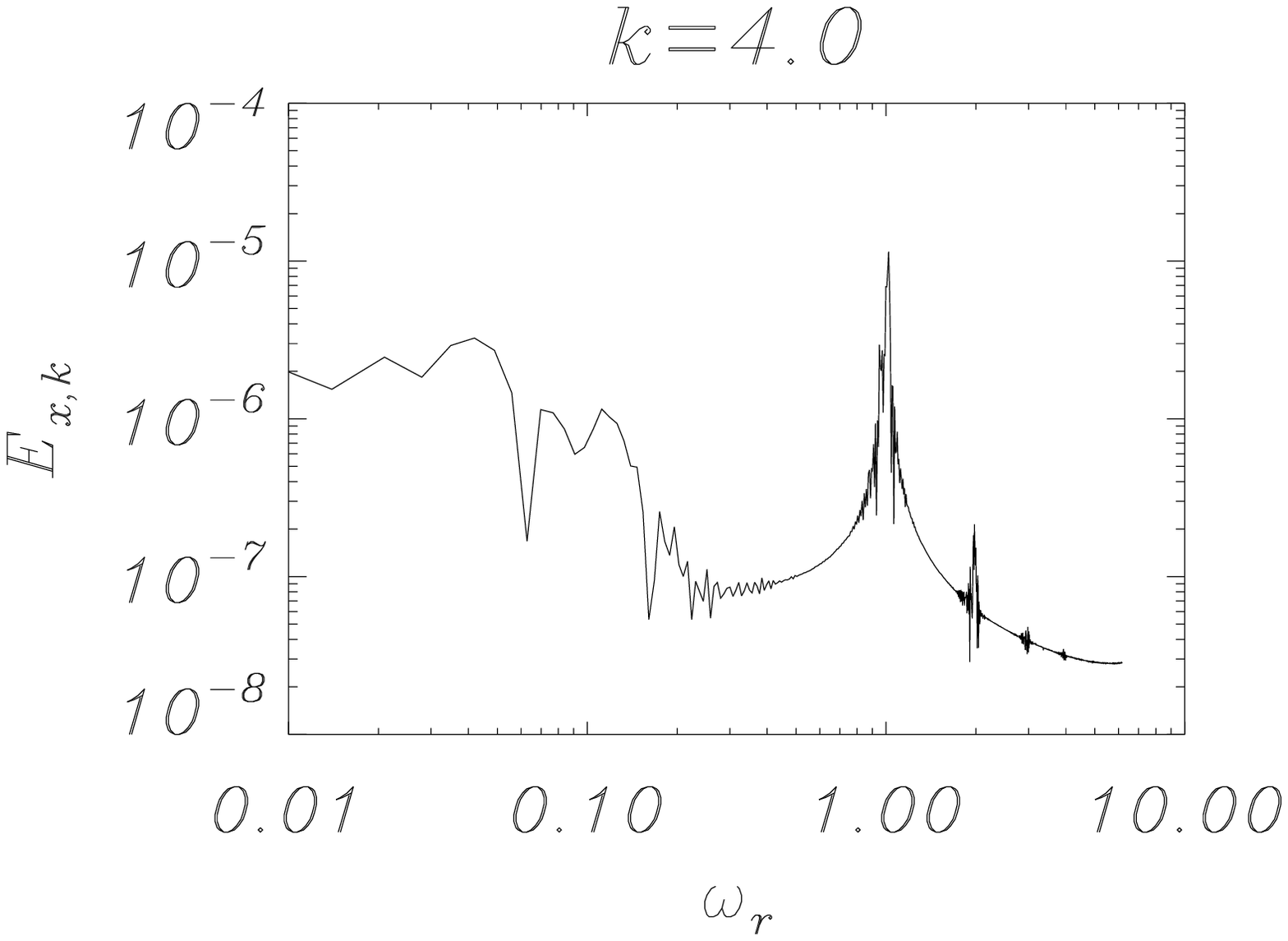,height=4.6cm,width=5.7cm}
\hspace{-0.4cm}\psfig{figure=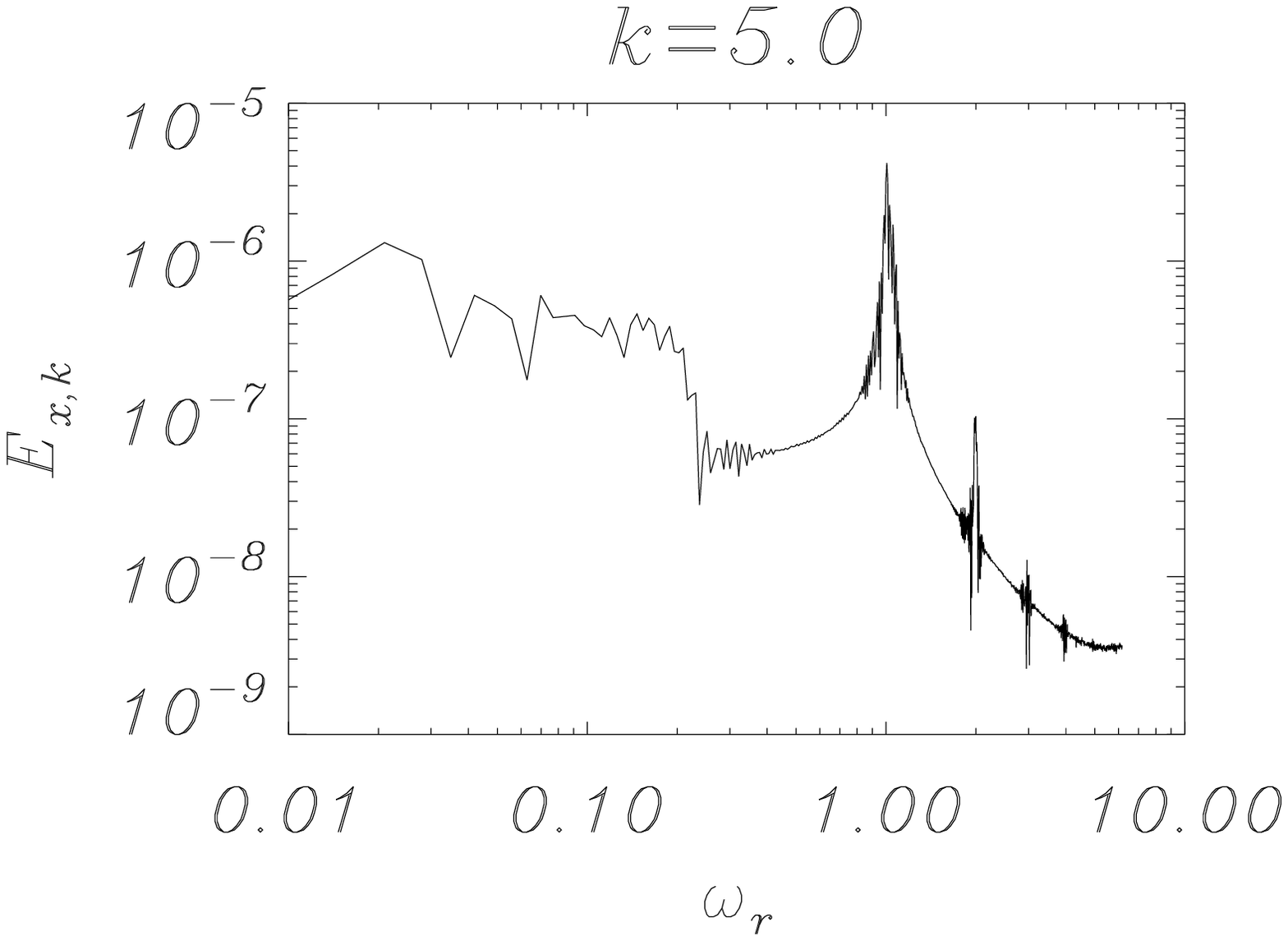,height=4.6cm,width=5.7cm}}
\centerline{\hspace{-0.2cm}\psfig{figure=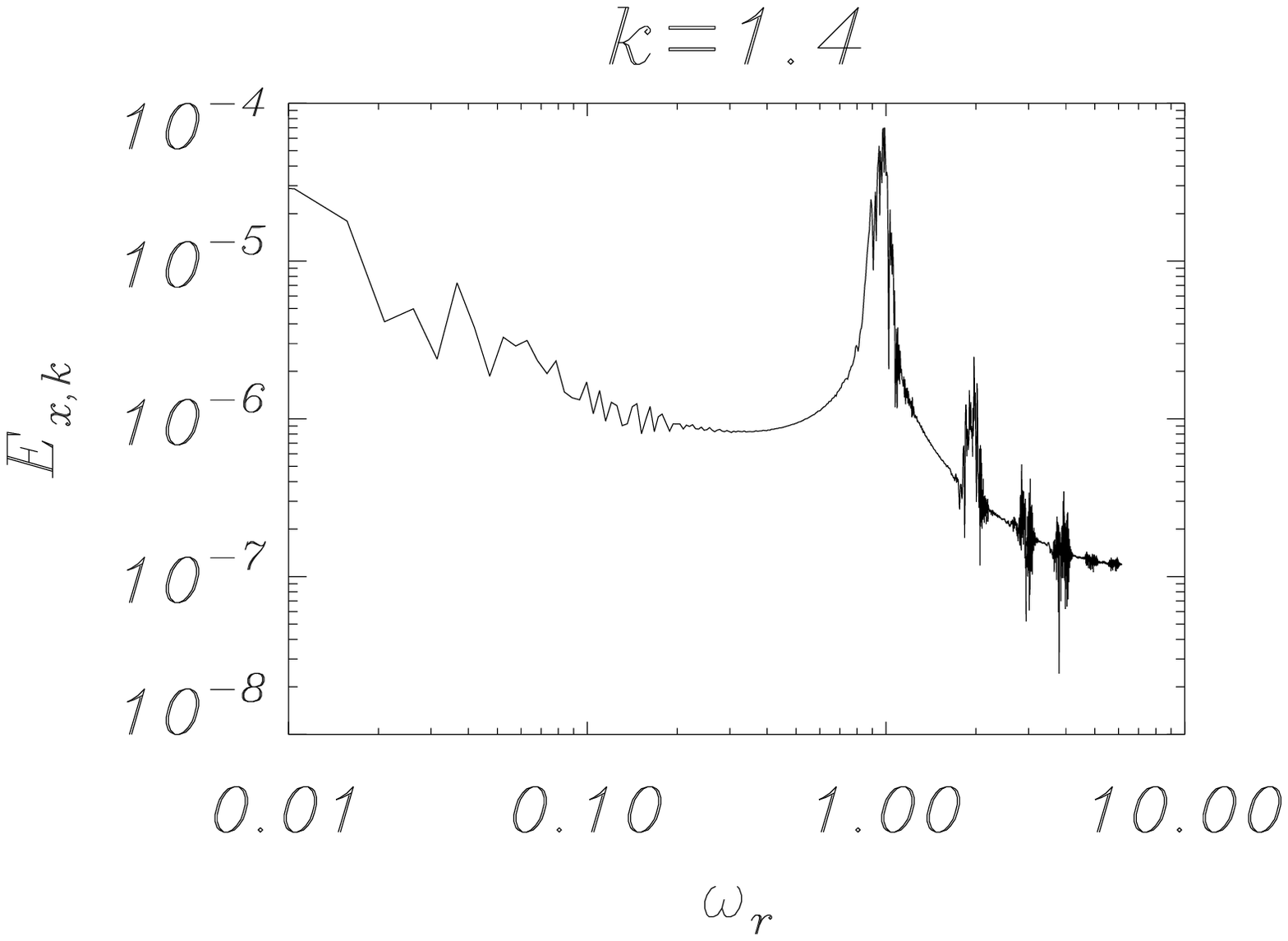,height=4.6cm,width=5.7cm}
\hspace{-0.4cm}\psfig{figure=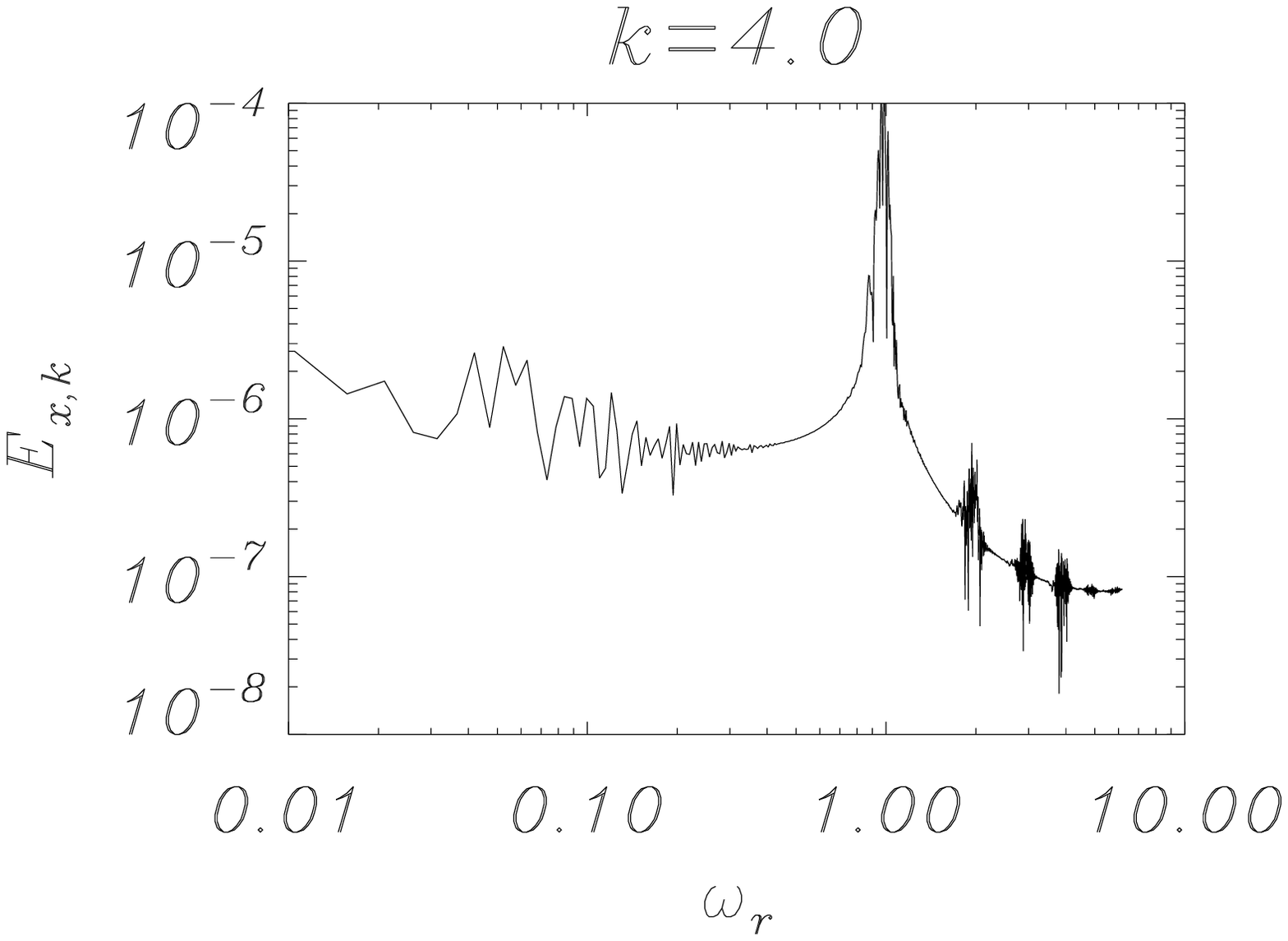,height=4.6cm,width=5.7cm}
\hspace{-0.4cm}\psfig{figure=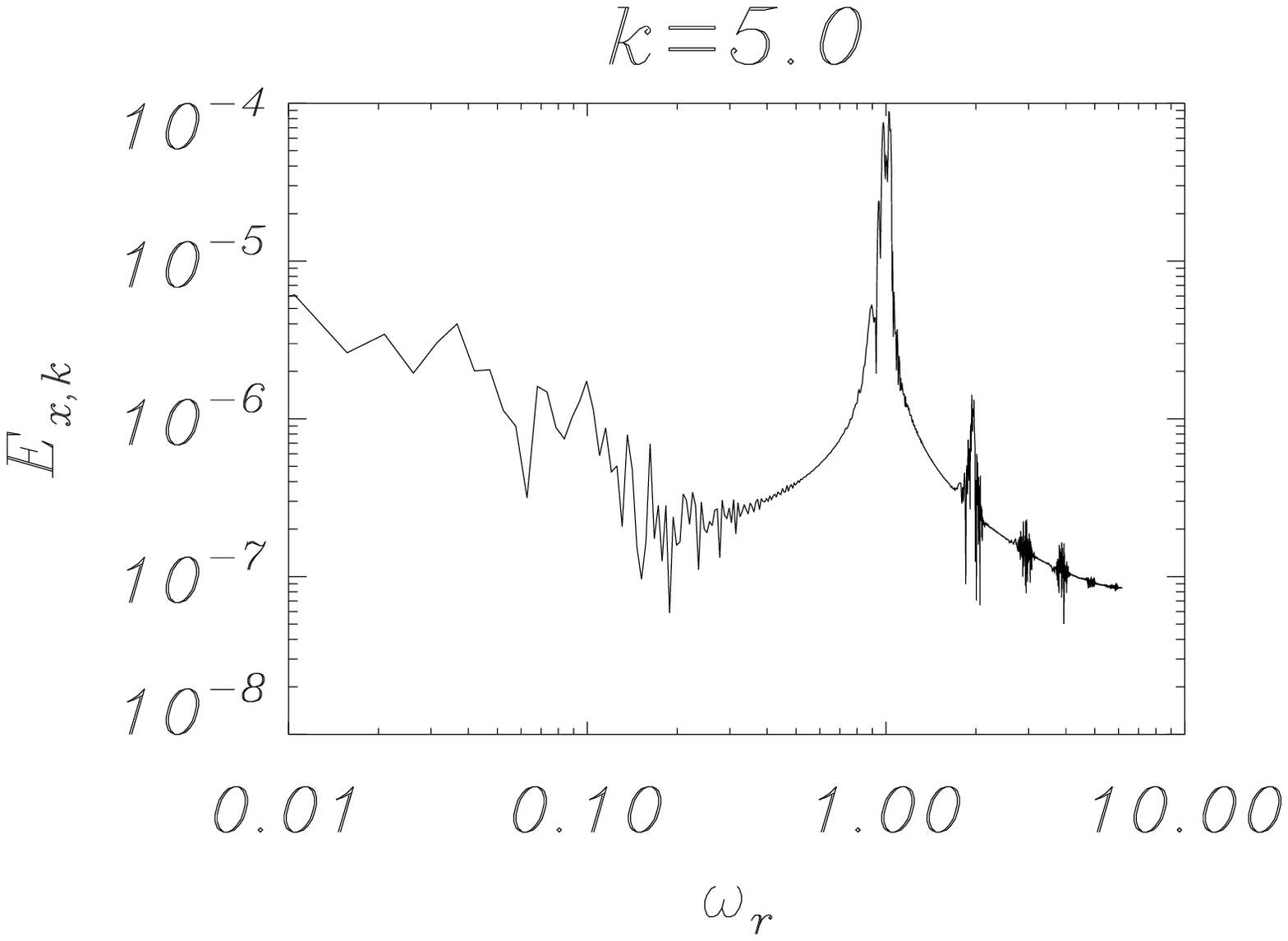,height=4.6cm,width=5.7cm}}
\vspace{-0.4cm}
\caption{\small Frequency spectrum of the Fourier component of the electrostatic field, $E_{x,k}$ at  $k = 1.4,4.0,5.0$ for $B_0 = 0$ (top frames) and $B_0 = 0.1$ (bottom frames).}
\label{Fig5}
\end{figure}
\begin{figure}[!h]
\vspace{-0.8cm}
\centerline{\hspace{-0.2cm}\psfig{figure=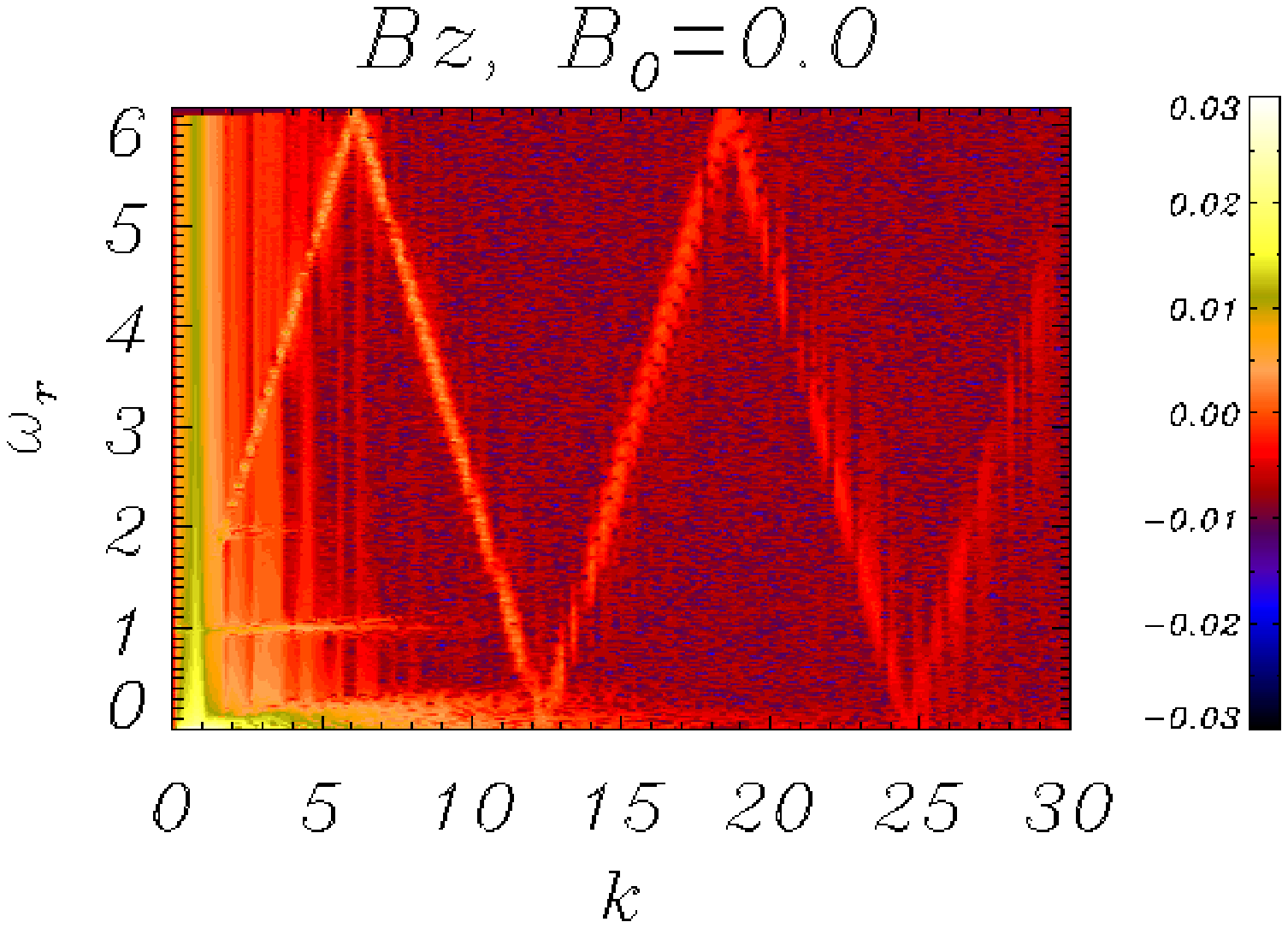,height=4.5cm,width=6.5cm}
\hspace{-0.6cm}\psfig{figure=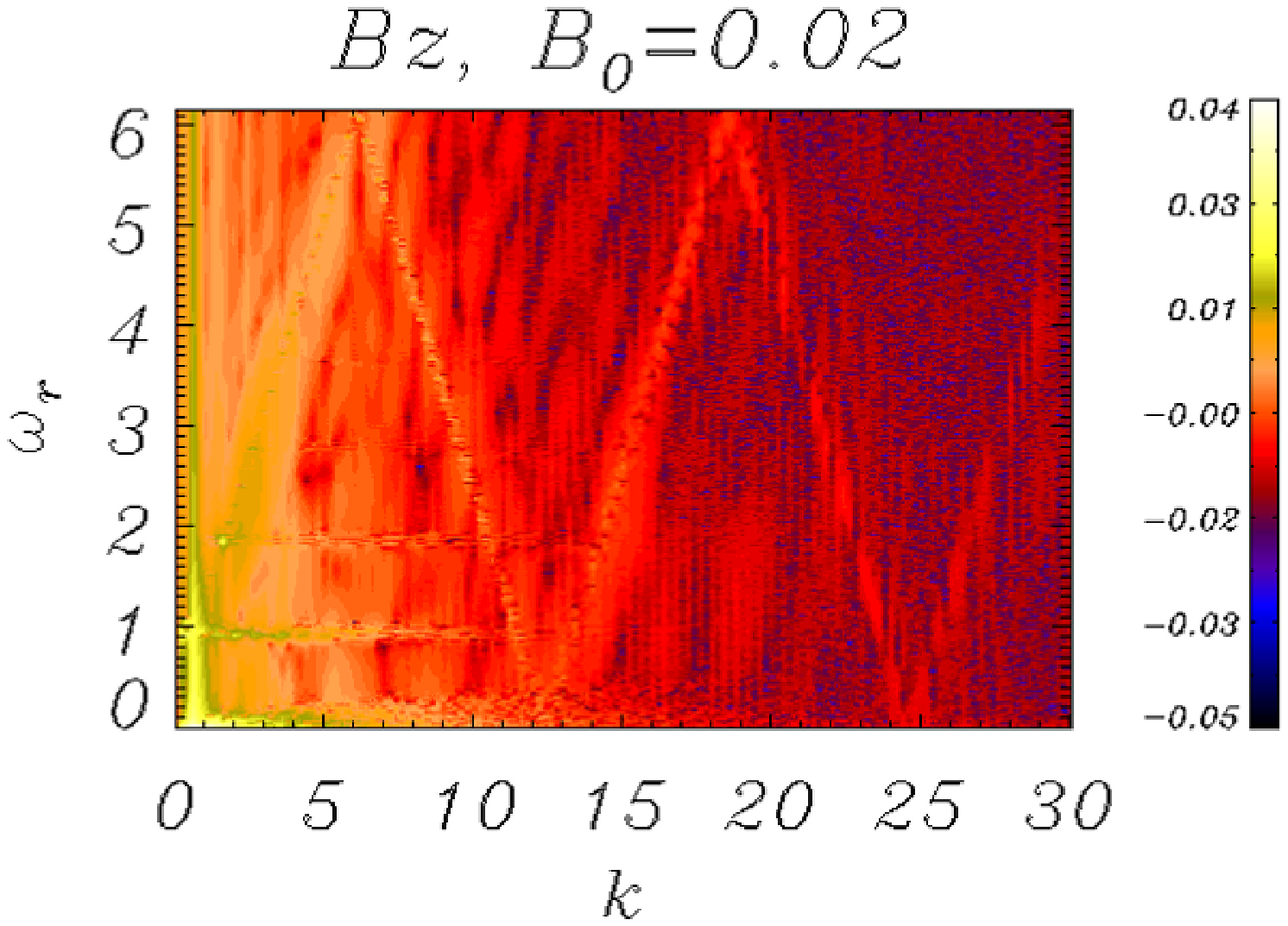,height=4.5cm,width=6.5cm}}
\vspace{-0.4cm}
\centerline{\hspace{-0.2cm}\psfig{figure=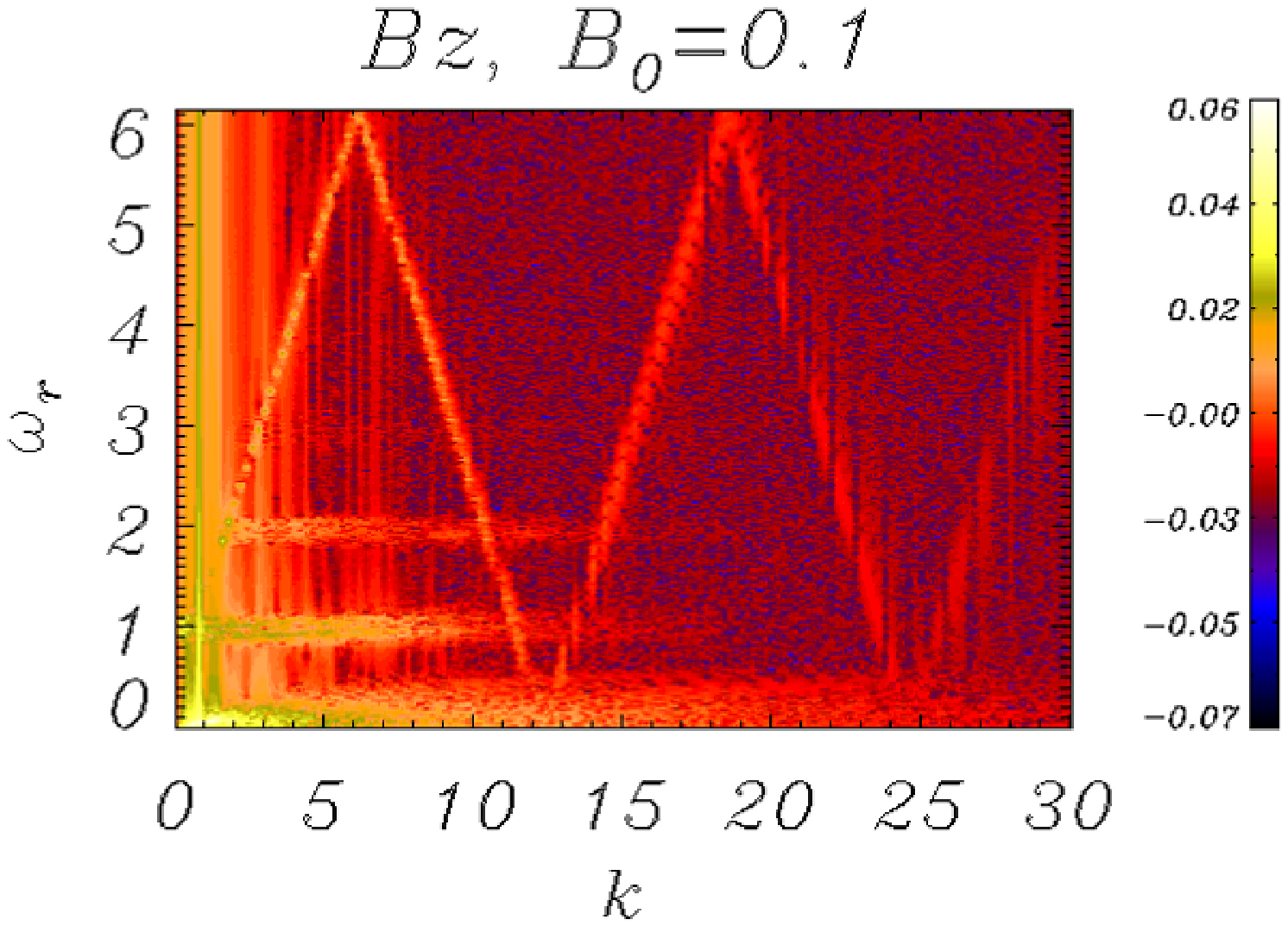,height=4.5cm,width=6.5cm}
\hspace{-0.6cm}\psfig{figure=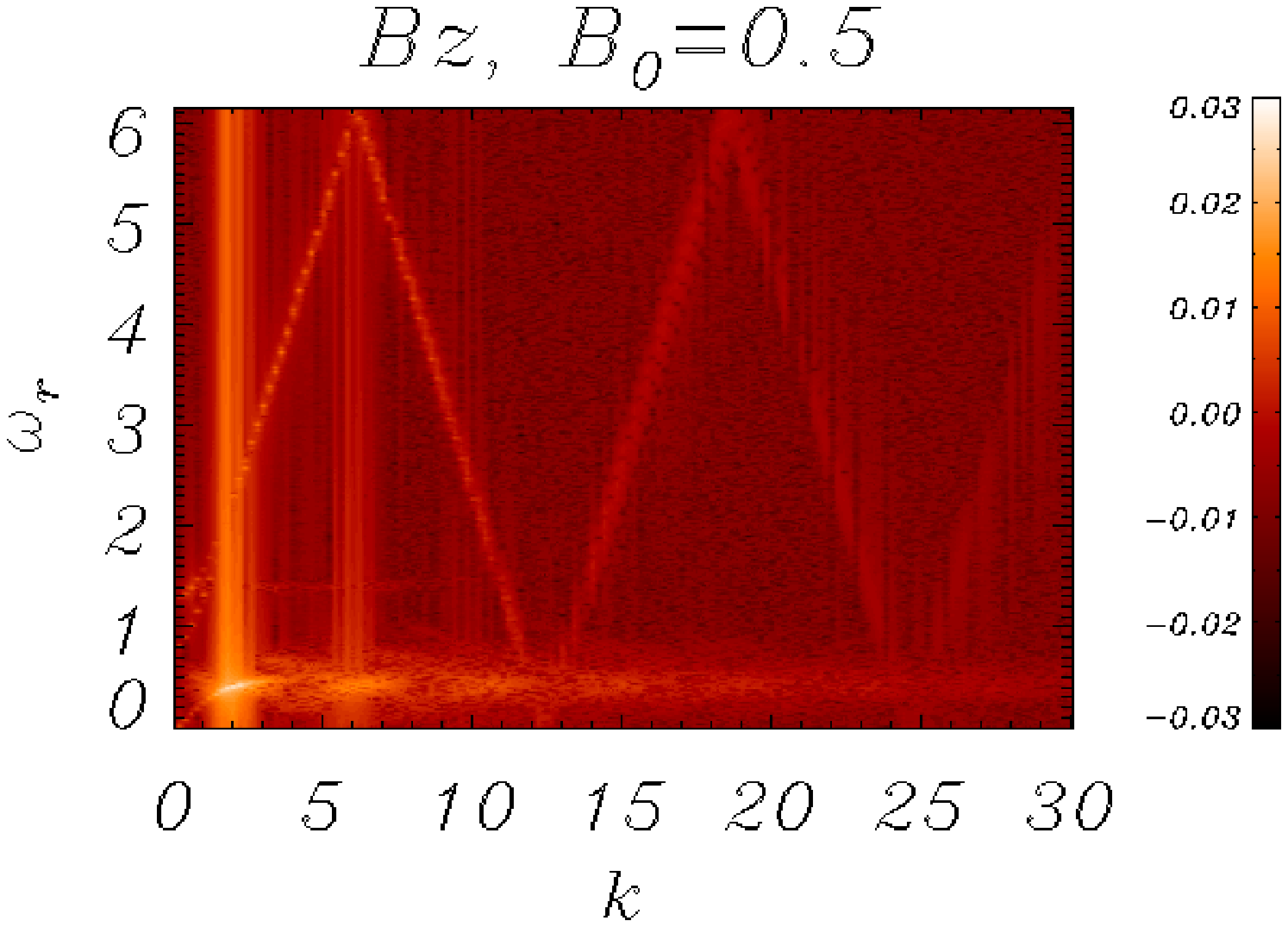,height=4.5cm,width=6.5cm}}
\vspace{-0.3cm}\caption{\small  Contour plot of the magnetic  field $B_z$ in  the  $\omega_r$ - $k$ space for $B_0 = 0, 0.02 , 0.1, 0.5$. Note the  finite time sampling  effect on the  frequency electromagnetic waves $\omega_r \sim k$ which leads to a ``zigzag''  pattern. Note the change in the scale of the color coding.}
\label{Fig4}
\end{figure}

\noindent
Electromagnetic  emission at  $3 \omega_{pe}$ has been reported in type II bursts \cite{Kliem1992} and in type III bursts \cite{Benz1973}.  Various theories have been put forward in order to explain this higher-harmonic emission. The coalescence of a Langmuir wave and  a $2\omega_{pe}$ electromagnetic  wave was suggested for the $3\omega_{pe}$  emission in Ref.\onlinecite{Zlotnik1978}.
 \\ Furthermore,   electromagnetic waves with a frequency $\omega_r \sim k$ can be seen in the spectra in Fig.\ref{Fig3} (see also bottom right frame in Fig.\ref{Fig1}): their frequency  does not correspond to harmonics of the plasma frequency and  shifts with $k$. The amplitudes  of these latter waves are smaller  than those of the  $\omega_{pe}$ waves,  but greater than those of the  $2\omega_{pe}$ and $3 \omega_{pe}$ emissions. The higher harmonics  emission and the  waves at $\omega_r \sim k$ are present only at higher wave mode numbers.
\\ Multiple harmonic emission up to fifth harmonic in the Earth's fore-shock environment has been reported (see Ref.\onlinecite{Cairns1986}). In our simulations the  relative intensity of the emission decreases rapidly with increasing harmonic number with harmonics greater than three being too faint to be observed.  Although in the magnetic field spectrum we do not observe electromagnetic emission at the fifth harmonic,  such high harmonics  are observed in the electrostatic spectrum as shown in Fig.\ref{Fig5}.
These $\omega_{pe}$ and higher harmonic  emissions are suppressed for larger  values of the ambient magnetic field. In a typical solar wind configuration the ratio $\omega_{ce}/\omega_{pe}$ is of the order\cite{Parks2003} of $~10^{-2}$. Hence only the choice of  ${\bf B}_0 = 0.02$ or $0.1$ represents a fairly typical situation for this kind of emission.  This can be clearly seen from  Fig.\ref{Fig4} where a contour plot of the perturbed magnetic field  $B_z$  in the  $\omega_r$-$k $ plane is shown.

\subsection{Electrostatic fields}\label {EF}

{The frequency spectra of the longitudinal field $E_{x,k}$ at $k= 1.4, 4, 5$ are shown in Fig.\ref{Fig5} for $B_0=0$ and for $B_0=0.1$. In both cases the spectra exhibit a clear peak at the electron plasma frequency. This peak for $B_0=0$ appears to be less pronounced for larger values of $k$, consistently with  what shown in Ref.\onlinecite{lopa1}, while for $B_0=0.1$ the amplitude  at $\omega_{pe}$ remains almost the same even at smaller wavelengths. In addition  in both cases we also see  peaks appearing at  multiple harmonics  of the plasma frequency. It should be noted though that for  the unmagnetized plasma case only harmonics up to $4 \omega_{pe}$ are observed in contrast to harmonics up to $5 \omega_{pe}$ that
are observed in the case with $B_0=0.1$. The amplitudes  of the multiple harmonics of $\omega_{pe}$ for $B_0=0.1$ are much larger  than in the $B_0=0$ case.}
{In addition, similarly to  the 1D-2V, $B_0=0$ analyzed in Ref.\onlinecite{lopa1}, electrostatic structures are formed corresponding to  plasma modes with values of $k$ larger than those of the whistler waves. For example, for $B_0=0$, the resonant velocity corresponding to the clear bump in the distribution function shown in Fig.\ref{Fig8} (top right frame) is $v_r=0.05$, leading to the excitation of Langmuir waves with wave mode $k \sim \omega_{r}/v_r  \sim 20$, while for $B_0=0.1$, the resonant mode at which the Langmuir waves are excited is $k \sim 10$}. As discussed in Ref.\onlinecite{lopa1}, these modes are excited by the deformation of the electron distribution function due to the differential rotation of the electrons in phase space in the magnetic field generated in the $y$-$z$ plane by the Weibel and by the whistler instabilities.
\begin{figure}[!h]
\centerline{\hspace{-0.2cm}\psfig{figure=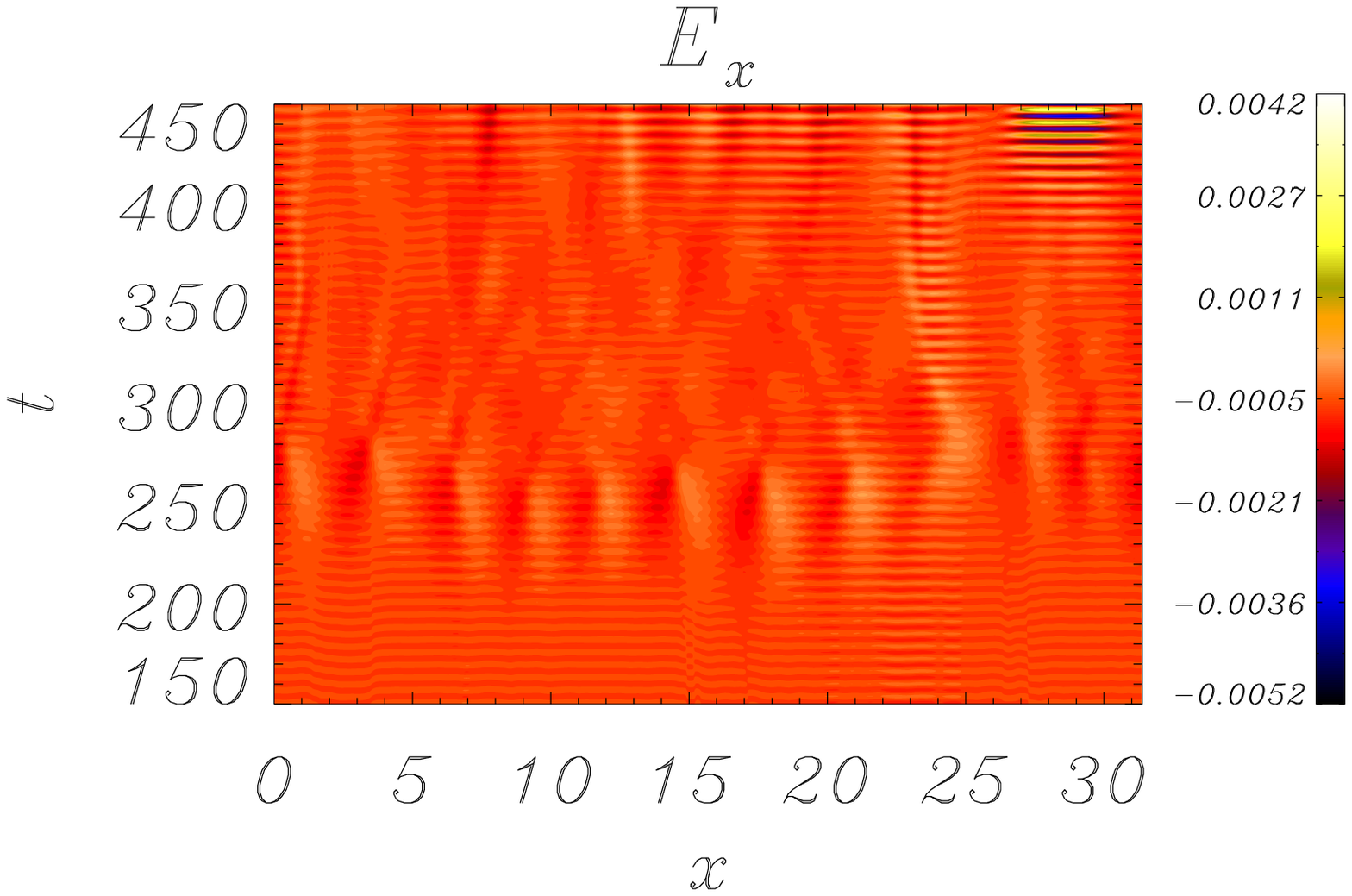,height=5.3cm,width=7.cm}
\hspace{-0.4cm}\psfig{figure=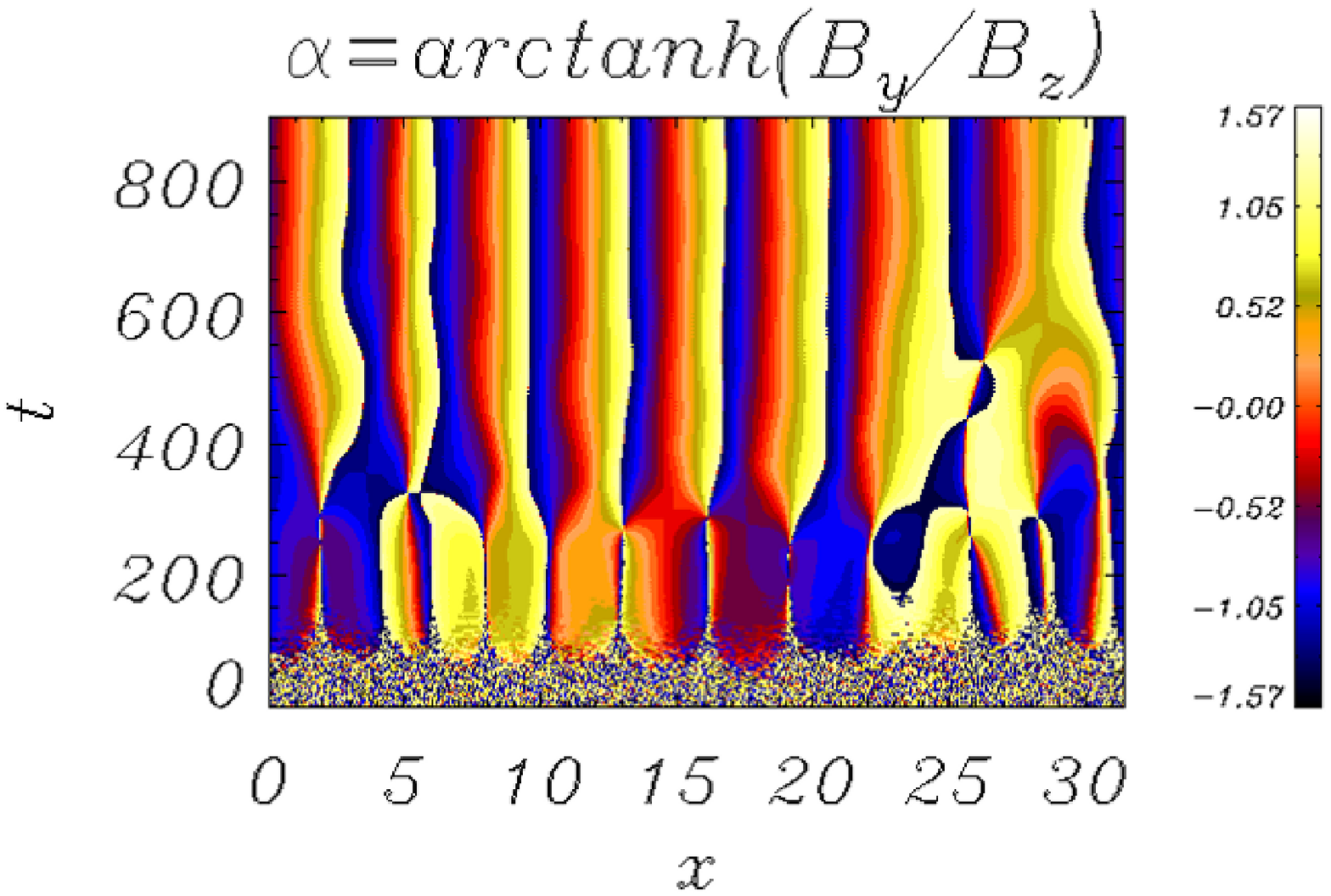,height=5.3cm,width=7.cm}}
\vspace{-0.4cm}
\caption{\small Left frame: contour plot of the electrostatic field in the $x$-$t$ plane (note only a  part of the time interval of the total time of the simulation is shown). Right frame: the angle $\alpha = arctanh(B_y/B_z)$  between the magnetic field components in $y$ and $z$ direction  in the  $x$-$t$ plane.}
\label{Fig6}
\end{figure}
\noindent This deformation becomes weaker as the effect of the ambient magnetic field $B_0$ is increased.  In addition if we compare the results obtained  for  $B_0=0$ in the  1D-2V case (see Fig.12  in Ref.\onlinecite{lopa1})  with the corresponding results shown in Fig.\ref{Fig6} (left frame)  for the 1D-3V case, we see that in the latter case  the amplitude  of these structures is somewhat smaller than in the 2V case. This difference  can be attributed to the fact that the 2V case imposes a  higher coherency of the perturbed magnetic field which is necessarily  oriented along the $z$ axis. On the contrary, in the 3V case  considered here,  the orientation of the perturbed magnetic field in the $y$-$z$ plane can change as a function of $x$.  The eventual formation of coherent electrostatic  structures can be attributed to the fact that, for sufficiently long times, the initially randomly  orientated   perturbed magnetic  field becomes fairly organized, as shown by Fig.\ref{Fig6} (right frame) where the angle $\alpha = arctanh({B_y}/{B_z})$ is shown as a function of space and time.  This  allows for  the formation of peaks in the electron distribution function, that are sufficiently coherent in space so as to excite plasma waves by inverse Landau damping, see Figs.\ref{Fig7} and \ref{Fig8}.
\begin{figure}[!h]
\vspace{-0.2cm}
\centerline{\hspace{3.8cm}\psfig{figure=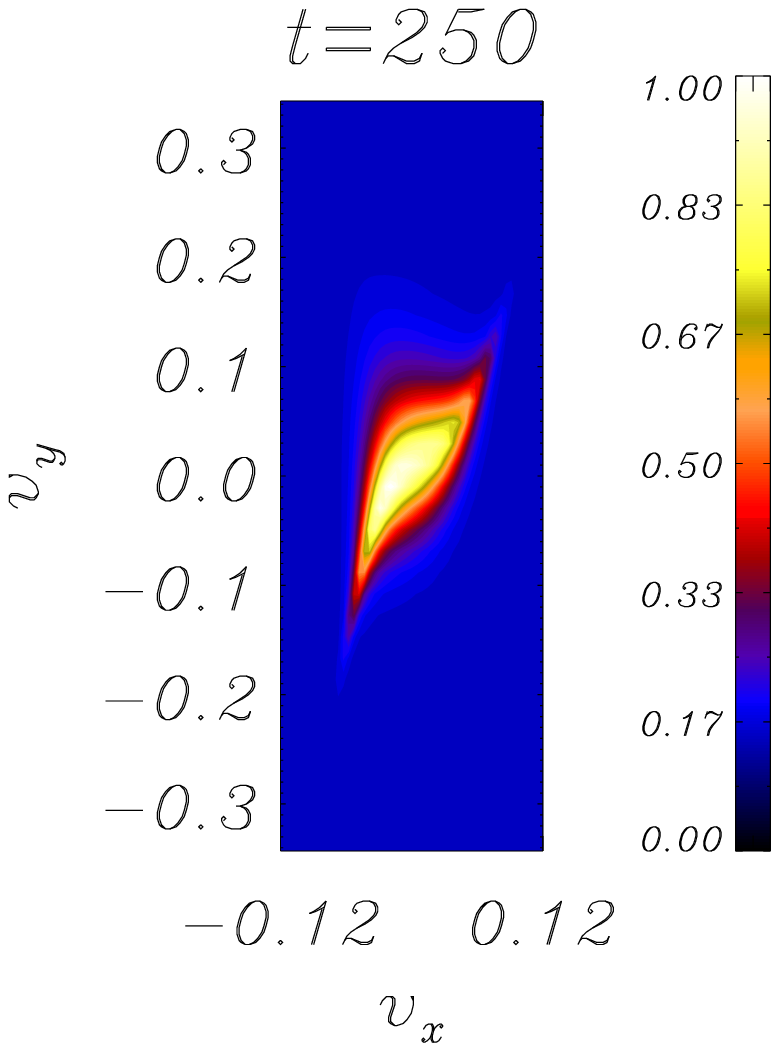,height=6.cm,width=8.5cm}
\hspace{-2.8cm}\psfig{figure=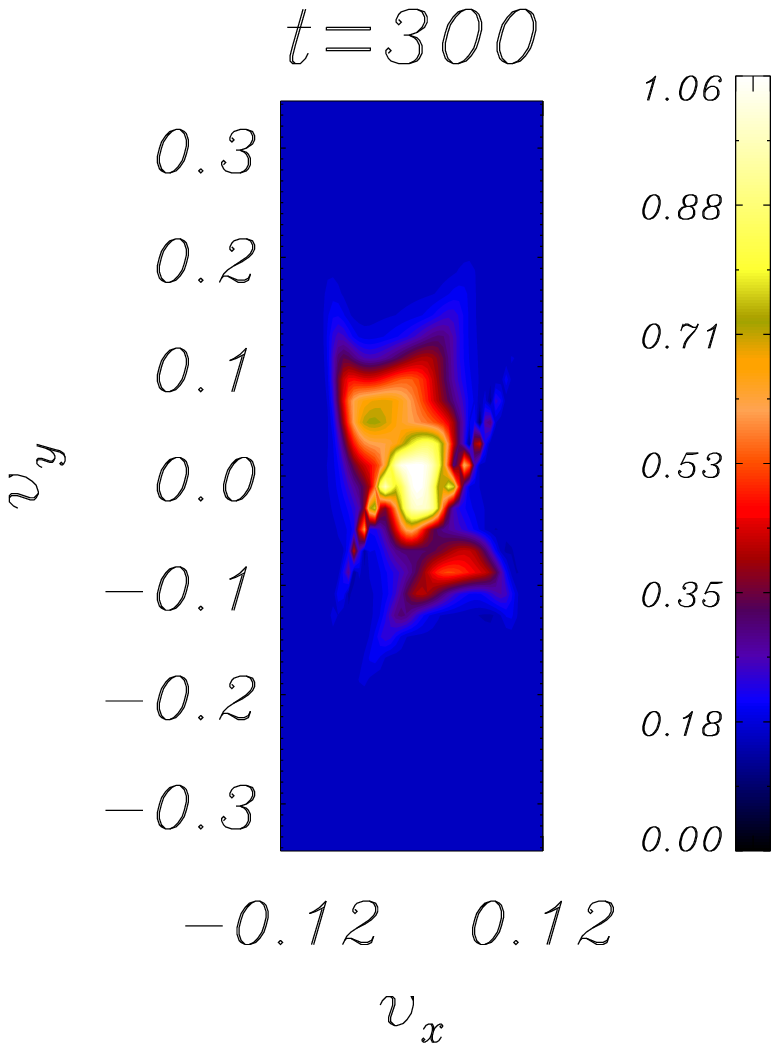,height=6.cm,width=8.5cm}
\hspace{-2.8cm}\psfig{figure=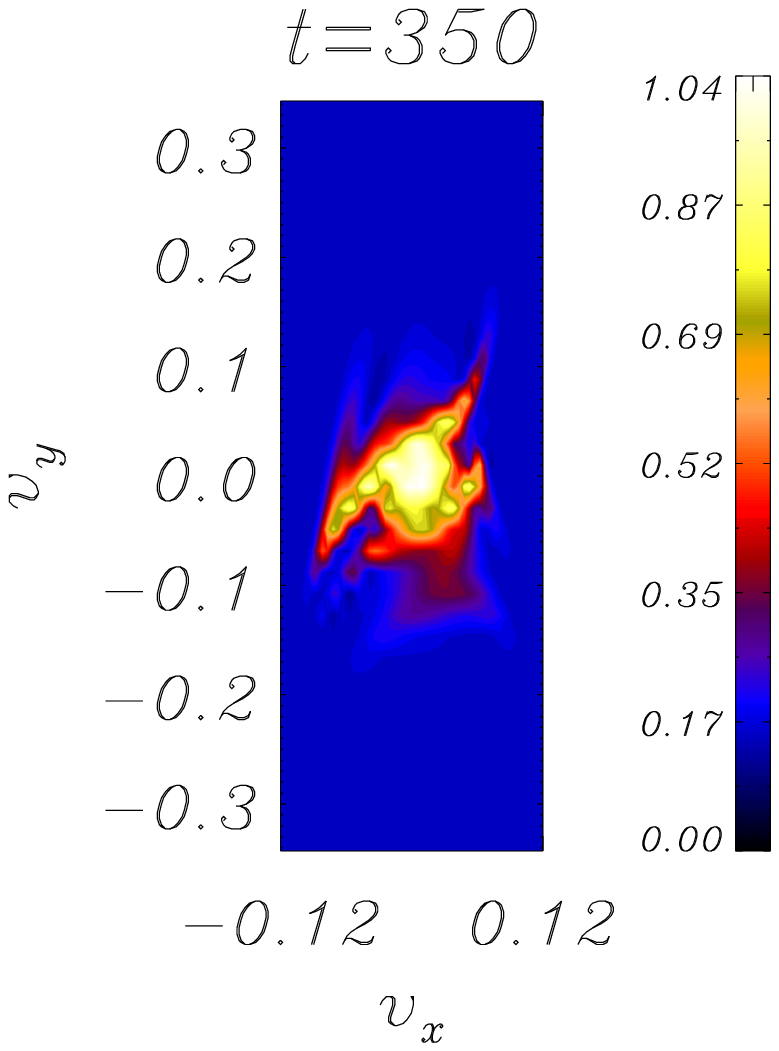,height=6.cm,width=8.5cm}}
\vspace{-0.8cm}
\centerline{\hspace{1.4cm}\psfig{figure=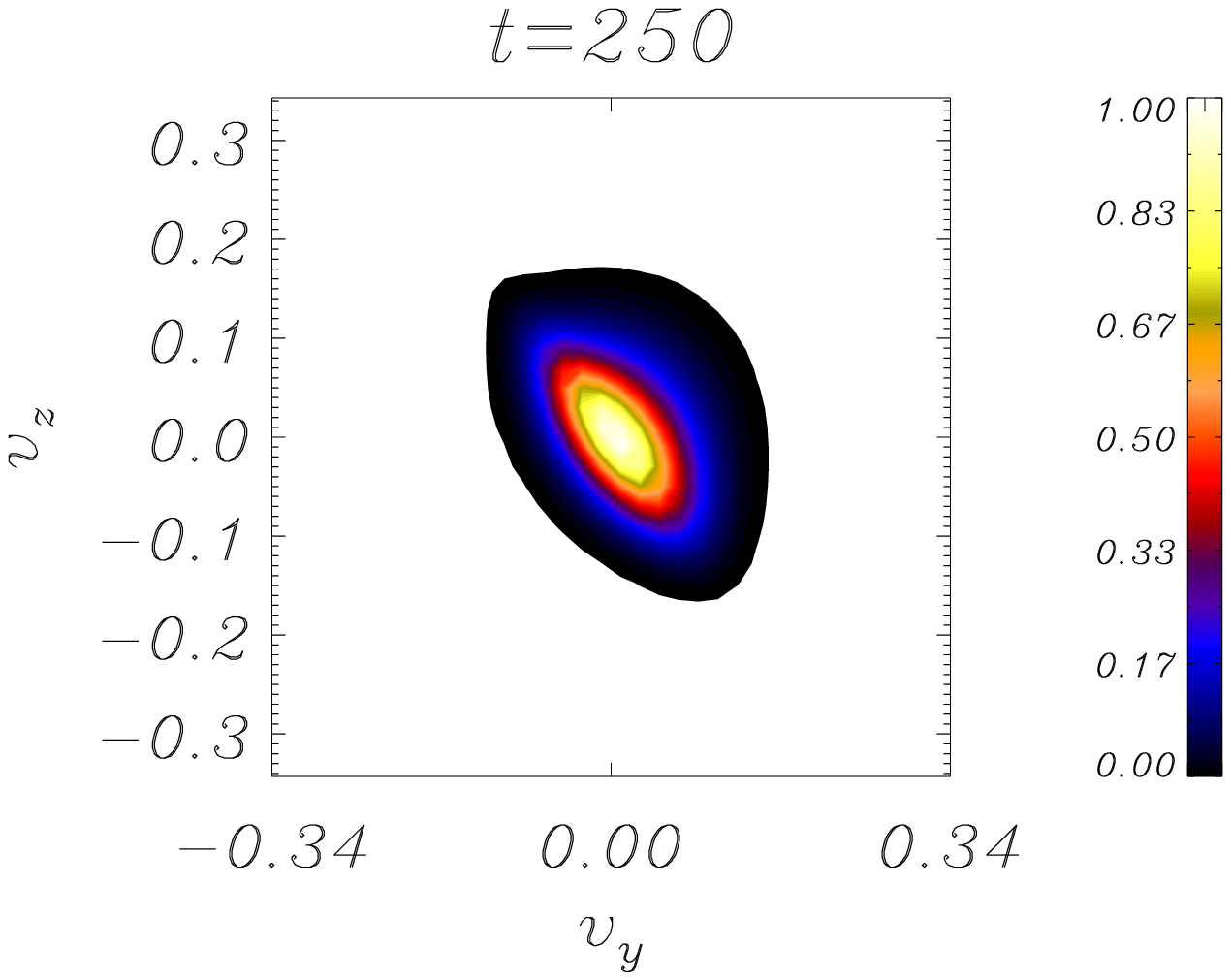,height=5.5cm,width=7.5cm}
\hspace{-1.8cm}\psfig{figure=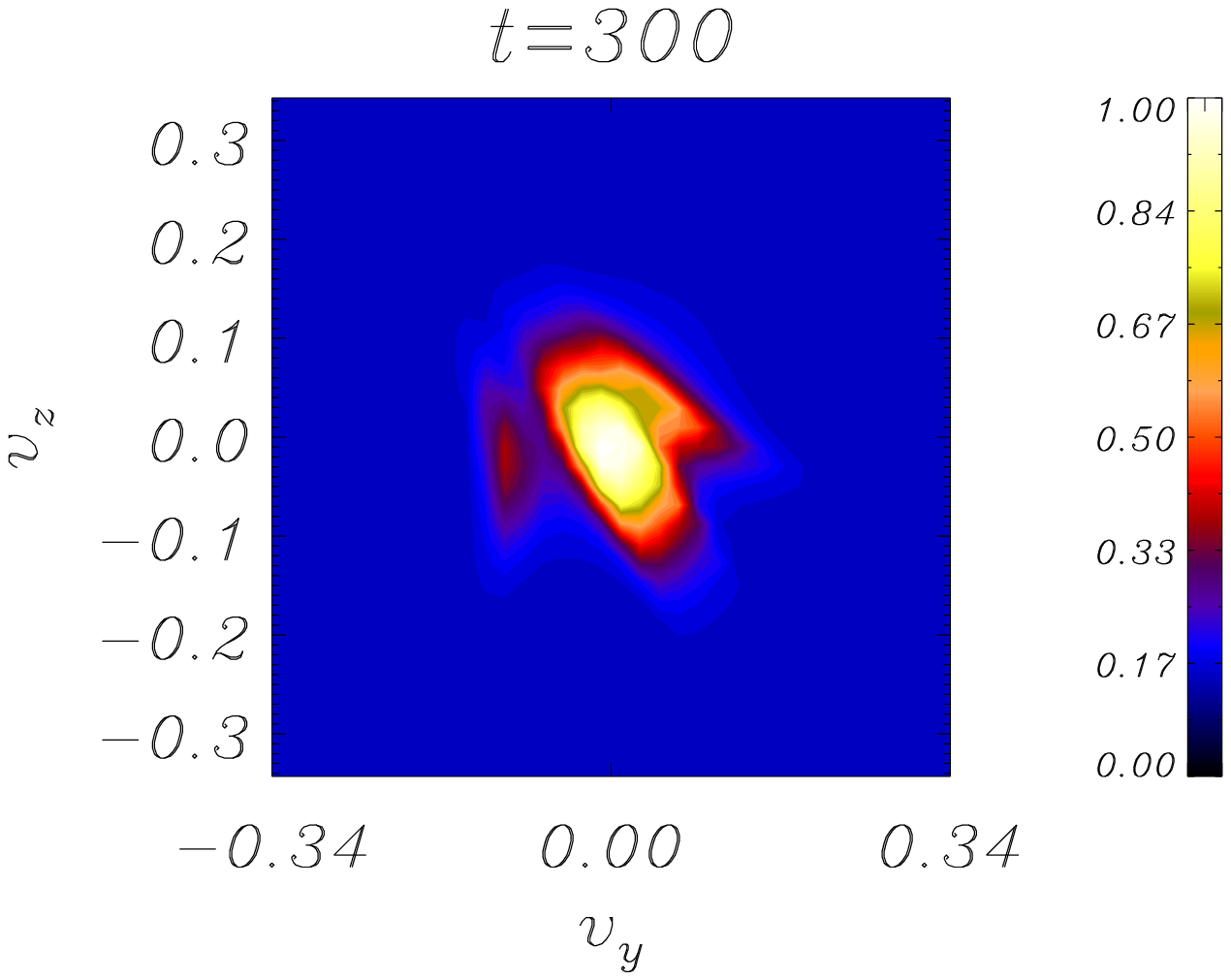,height=5.5cm,width=7.5cm}
\hspace{-1.8cm}\psfig{figure=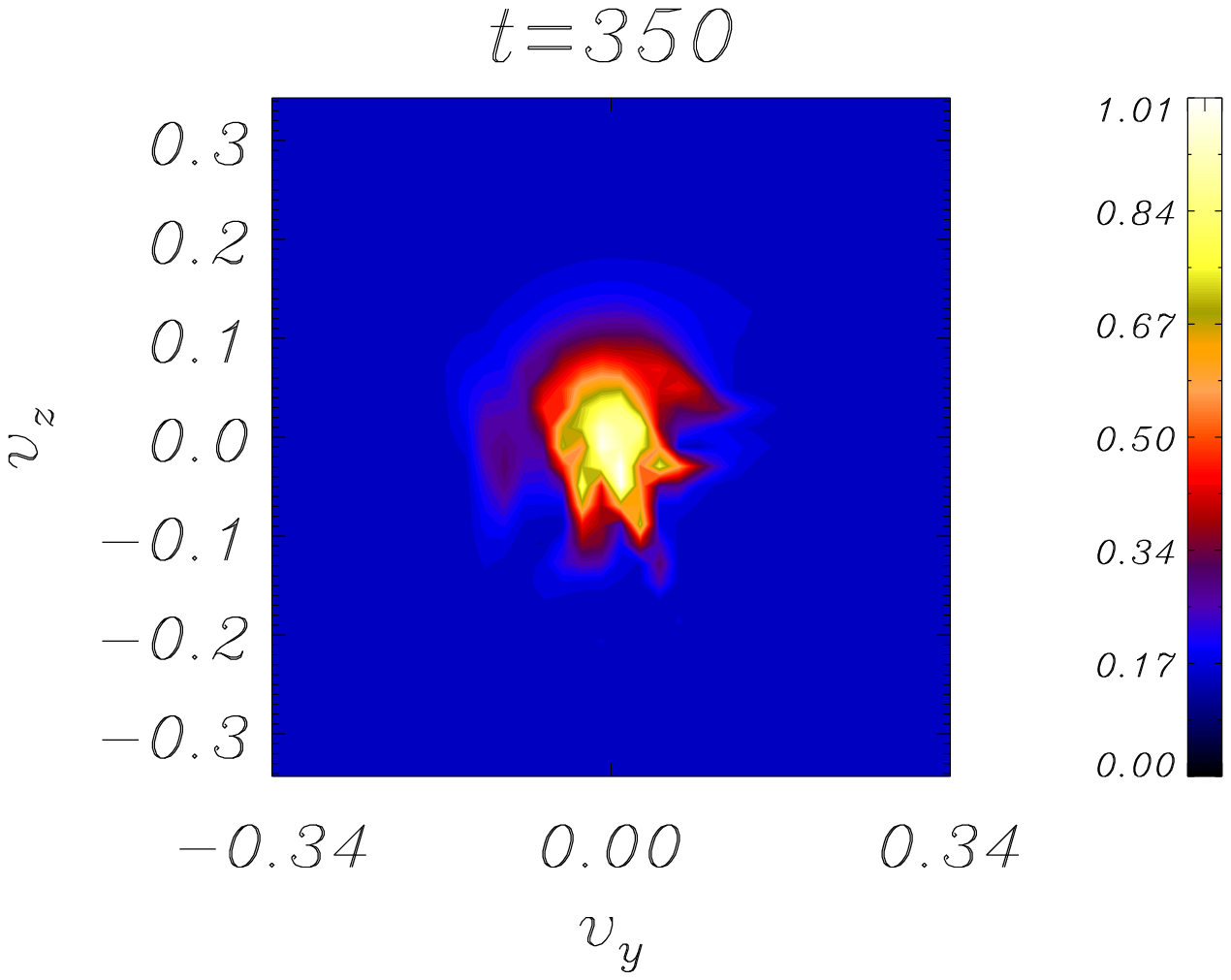,height=5.5cm,width=7.5cm}}
\vspace{-0.4cm}\caption[]{\small Top frames: Contour plot of the electron distribution function in the $v_x$-$v_y$ plane (for $v_z= 0$) for $B_0=0$ at $x=1.86$ (position corresponding to the  maximum value of the $z$ component of the perturbed magnetic field) for $t=250$, $t=300$, $t=350$. Bottom frames: Contour plot of the distribution function in the $v_y$-$v_z$ plane (for $v_x=0$) at $x=1.86$. Note the different velocity ranges. }
\label{Fig7}
\end{figure}
 {The contour plot of the electron distribution function in the $v_x $-$ v_y$ plane for $B_0=0$ is shown in Fig.\ref{Fig7}. The distribution function starts  being distorted,  with a differential  rotation combined with a spreading along $v_x$, at the time ($t=300$) when the instability begins to saturate. These deformations become 'multi-armed' as time evolves. Similar kind of deformations are observed in the $v_x $-$ v_z$ plane (not shown here). Multi-armed structures are also noticeable in the  $v_y$-$ v_z$ plane (Fig.\ref{Fig7}, bottom frames).
\begin{figure}[!h]
\vspace{-0.2cm}
\centerline{\hspace{-0.cm}\psfig{figure=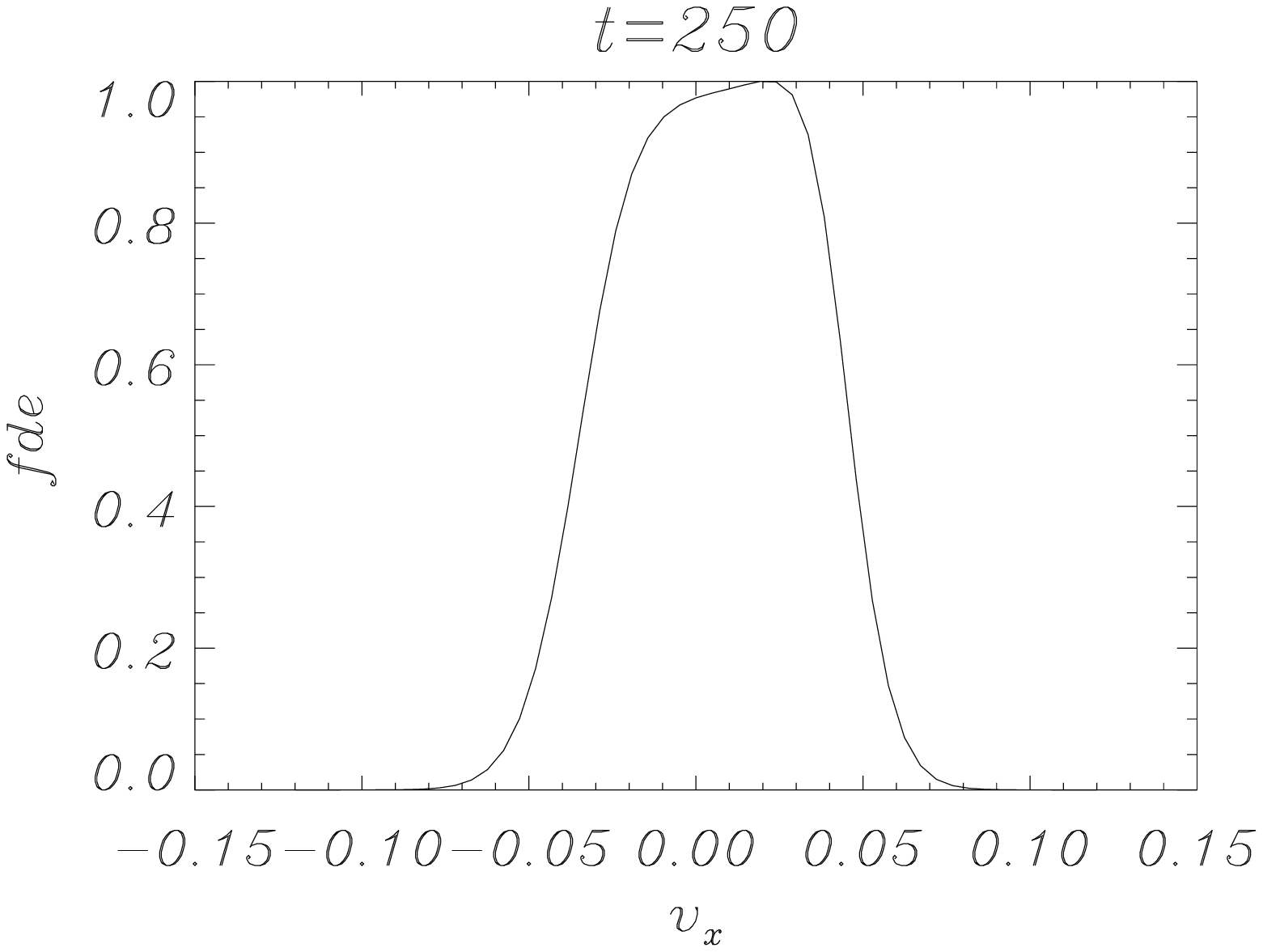,height=4.cm,width=6.cm}
\hspace{-0.cm}\psfig{figure=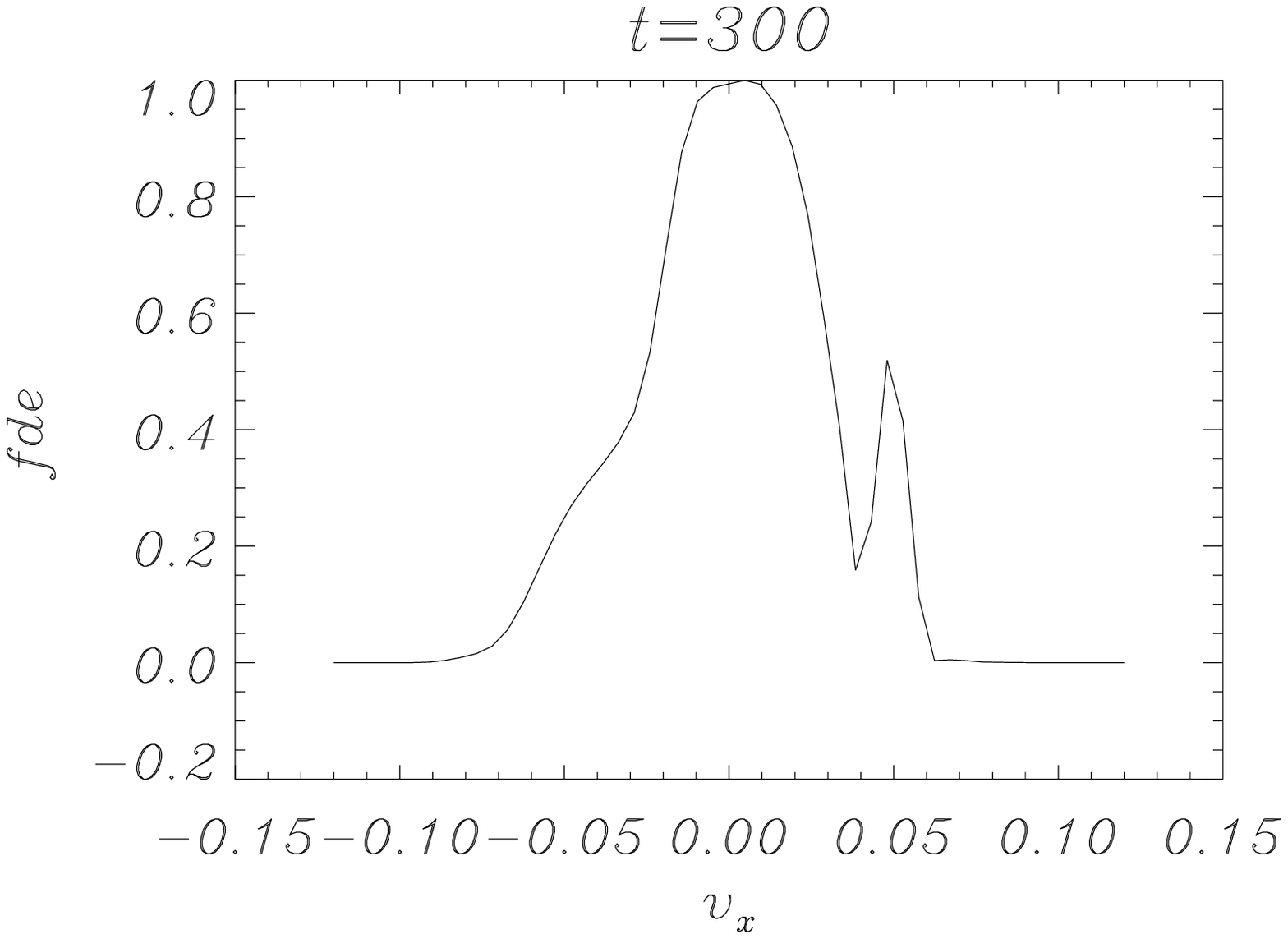,height=4.cm,width=6.cm}}
\centerline{\hspace{-0.2cm}\psfig{figure=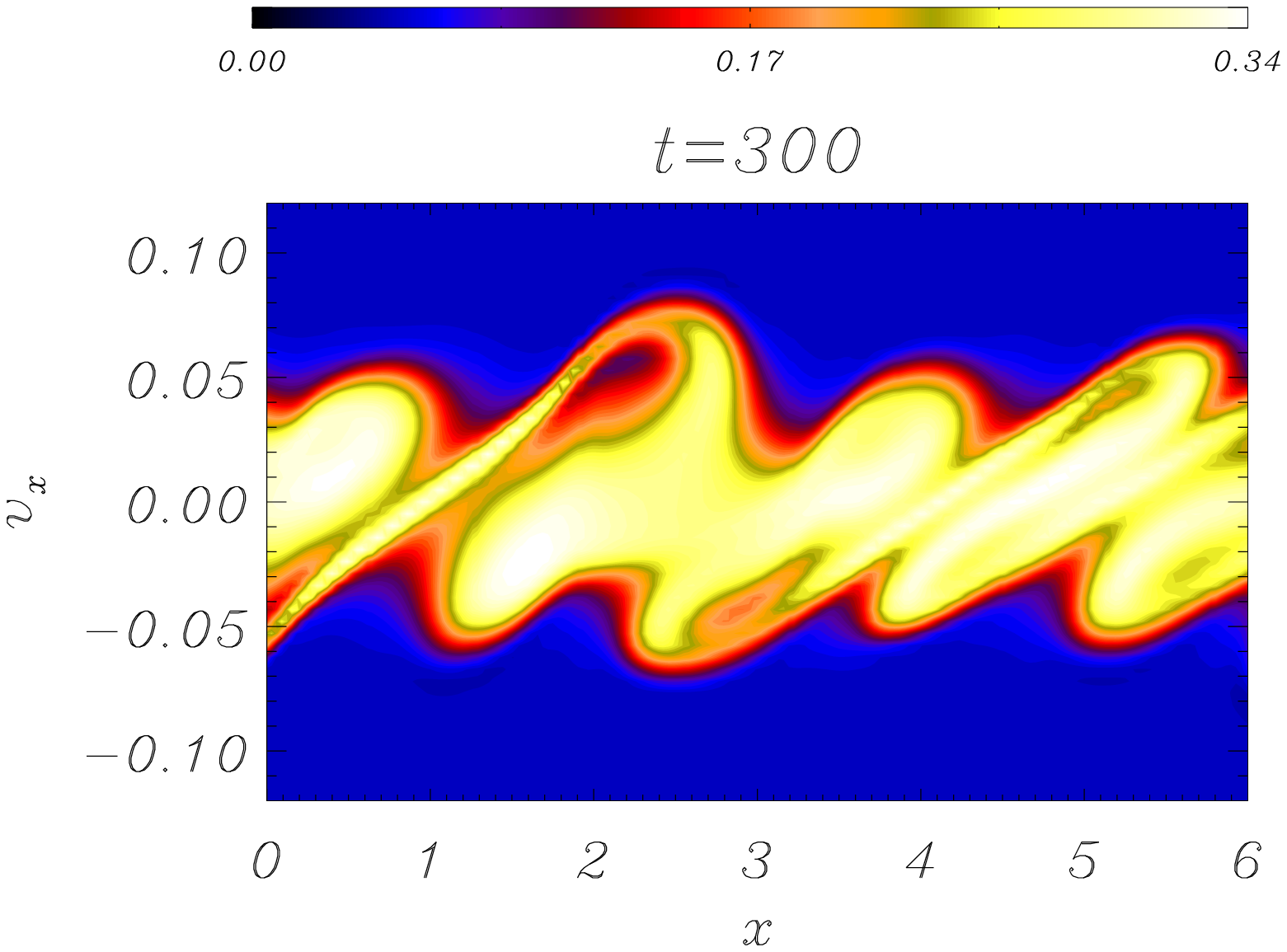,height=5.cm,width=7.cm}
\hspace{-1.cm}\psfig{figure=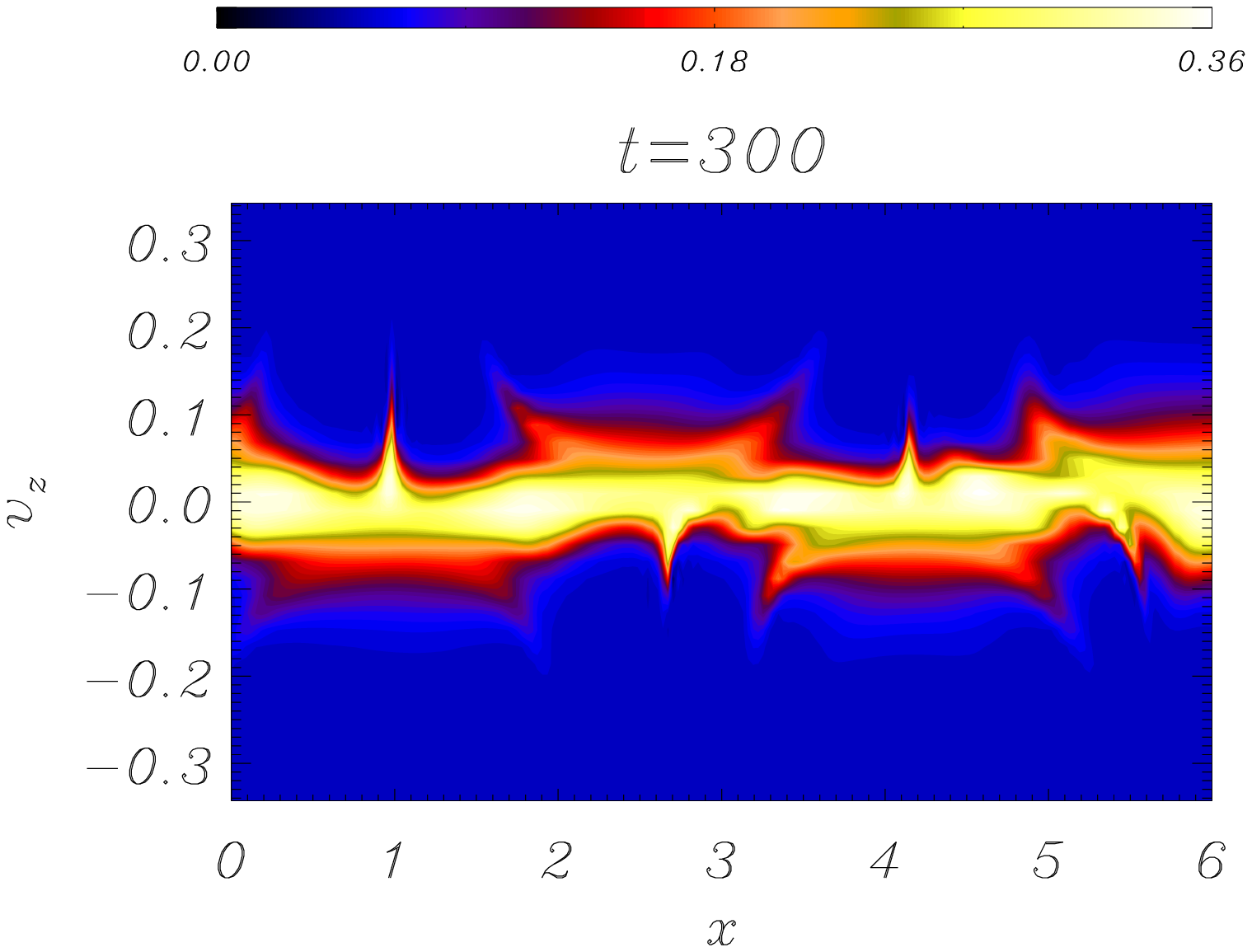,height=5.cm,width=7.cm}}
\vspace{-0.4cm}\caption[]{\small  Top frames: The normalized electron distribution function versus $v_x$ at $x=1.86$ for  $v_y=0$  and $v_z=0$ for $B_0=0$. Bottom (left) frame: Contour plot of the distribution function in the $x$-$v_x$ plane at  $v_y=0$ and $v_z=0$ and $x=1.86$. Bottom (right) frame: Contour plot of the distribution function in the $x$-$v_z$ plane at  $v_x=0$ and $v_y=0$ and $x=1.86$.}
\label{Fig8}
\end{figure}
The combined velocity and space dependence of the electron distribution function is shown in Fig.\ref{Fig8}, bottom frames.
The left frame shows the formation of phase space vortices in the $x$-$v_x$ plane (shown at  $v_y=0$ and $v_z=0$). The right frame shows  alternated filamented structures in the $x$-$v_z$ plane. The distribution function for $B_0=0.1$ behaves similarly (not shown here), but the strength of the deformations decreases.}

 \subsection{Low frequency modulations}

\begin{figure}[!h]
\vspace{-0.8cm}
\centerline{\hspace{-0.2cm}\psfig{figure=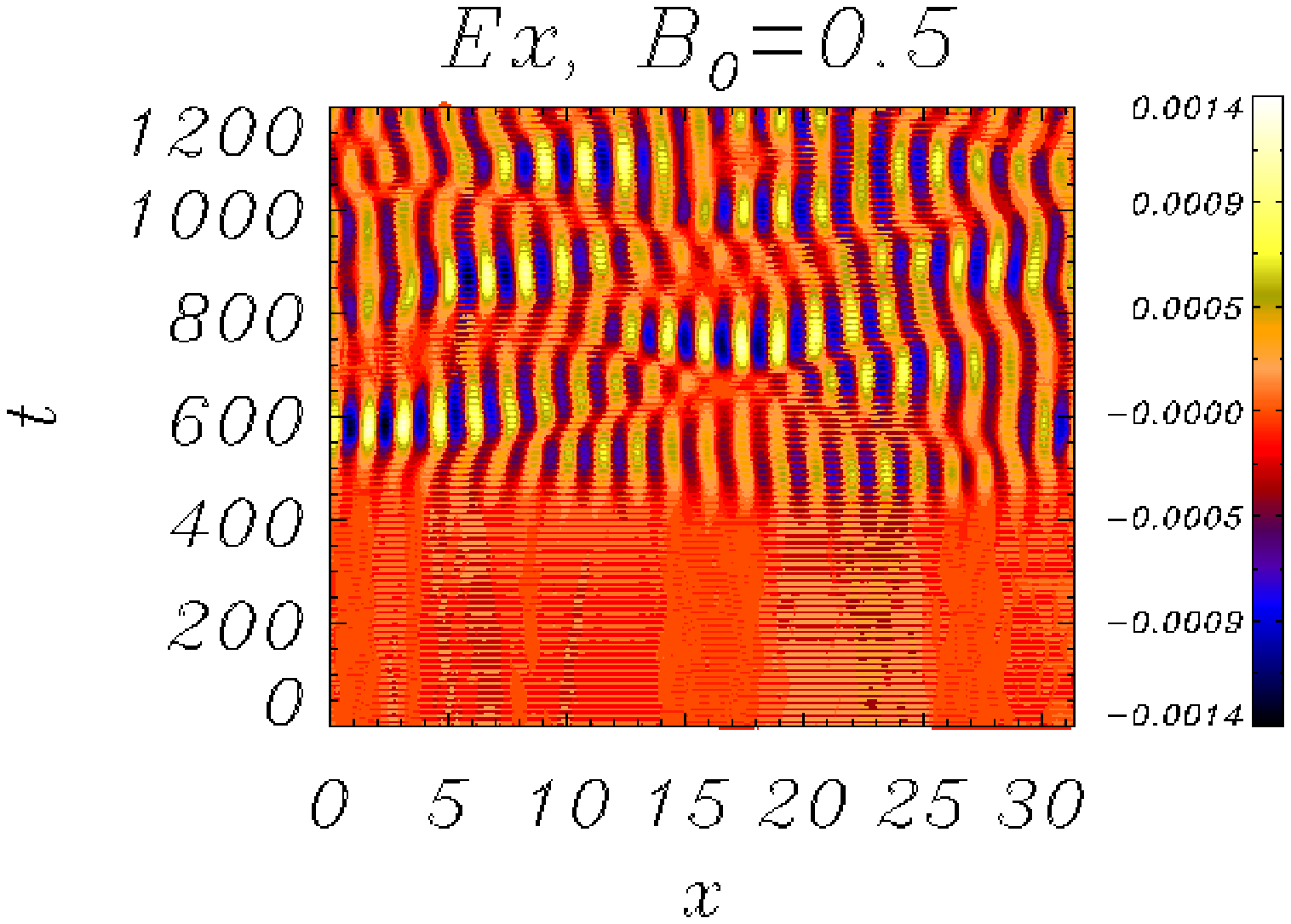,height=5.3cm,width=8.6cm}
\hspace{-0.6cm}\psfig{figure=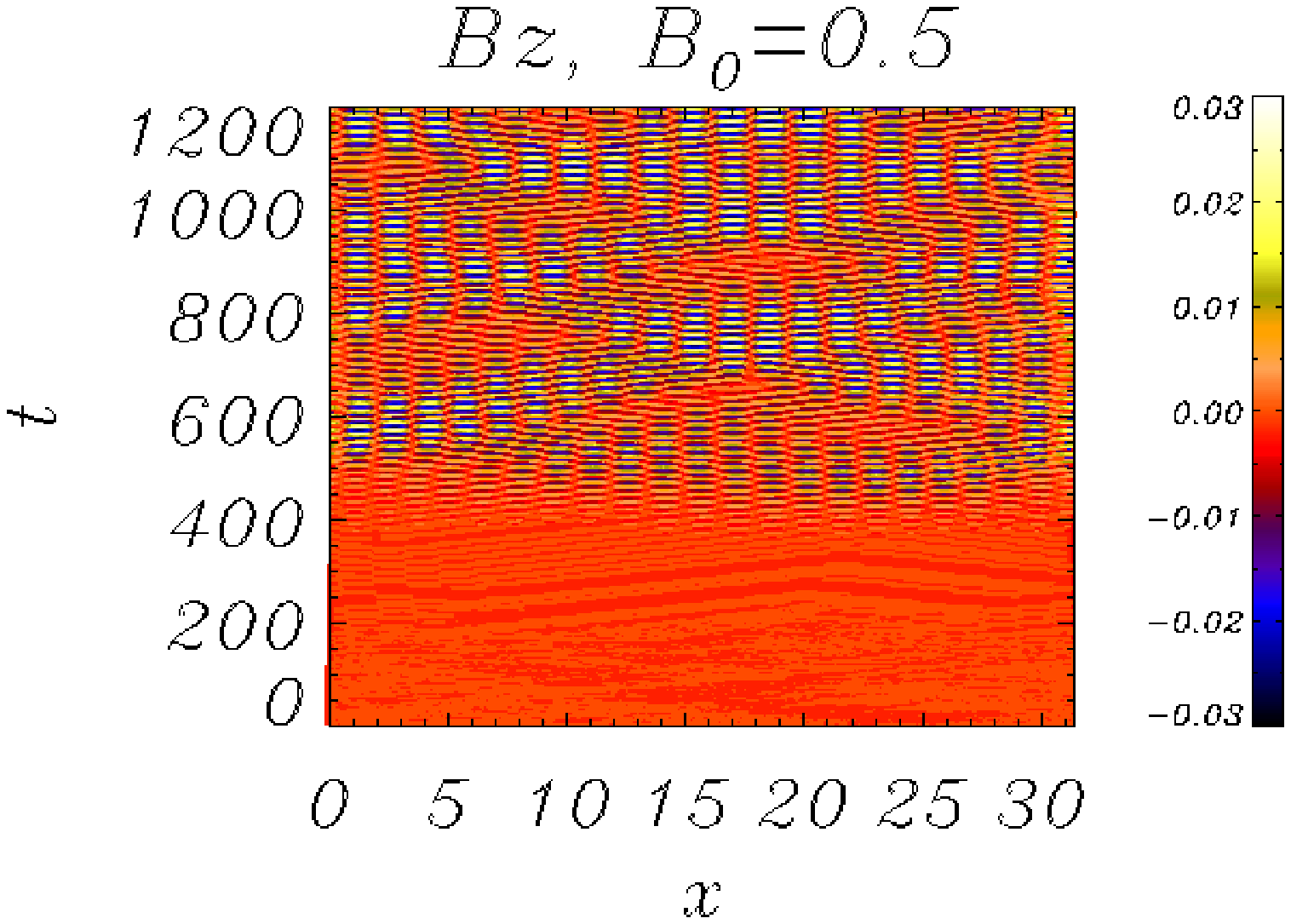,height=5.3cm,width=8.6cm}}
\vspace{-0.4cm}\caption{\small Contour plot of the electrostatic field (left frame) and of the magnetic field (right frame) in the $x $-$t$ plane for $B_0 = 0.5$.}
\label{Fig9}
\end{figure}

\begin{figure}[!h]
\vspace{-0.0cm}
\centerline{\hspace{-0.3cm}\psfig{figure=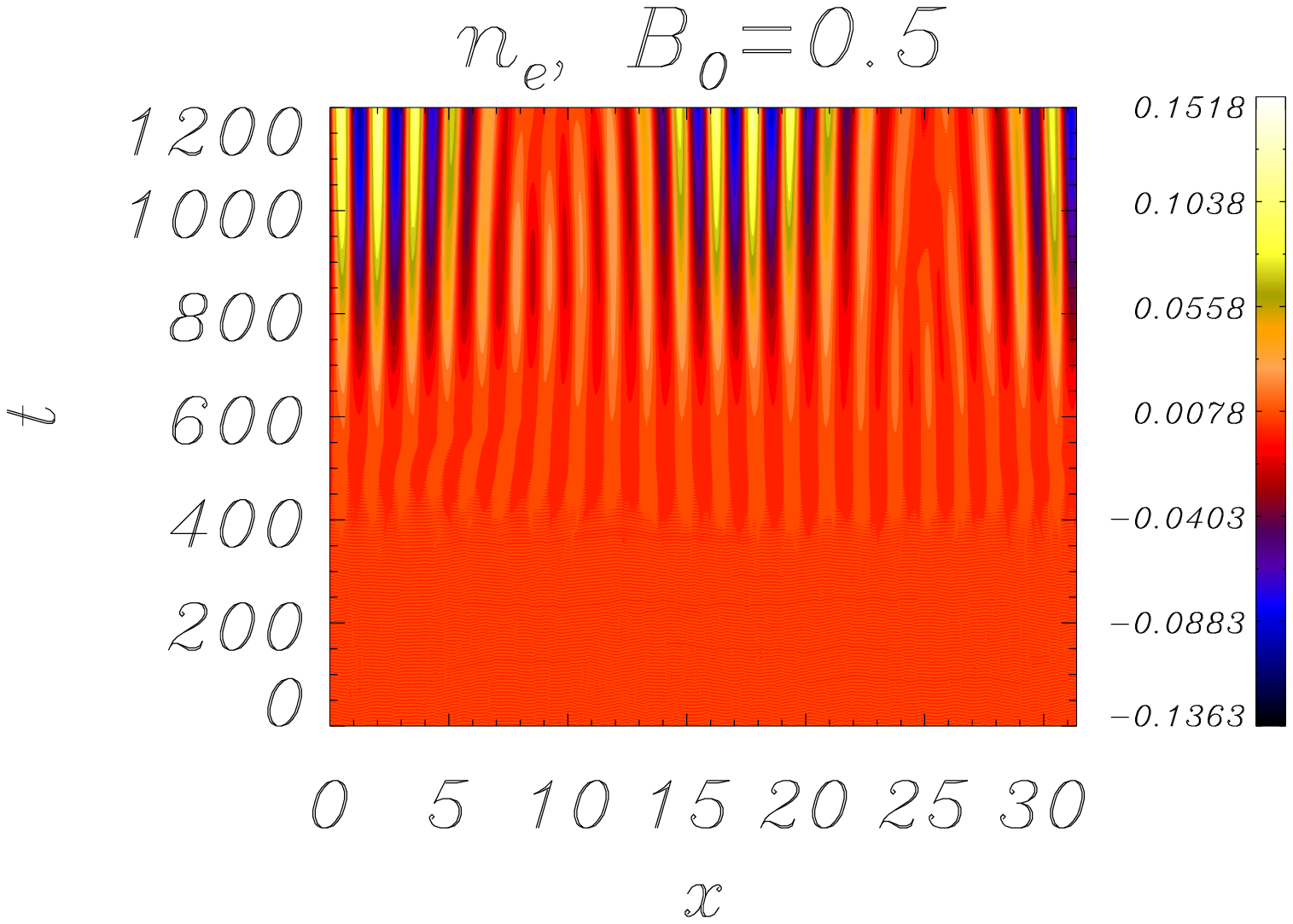,height=5.3cm,width=8.6cm}
\hspace{-0.3cm}\psfig{figure=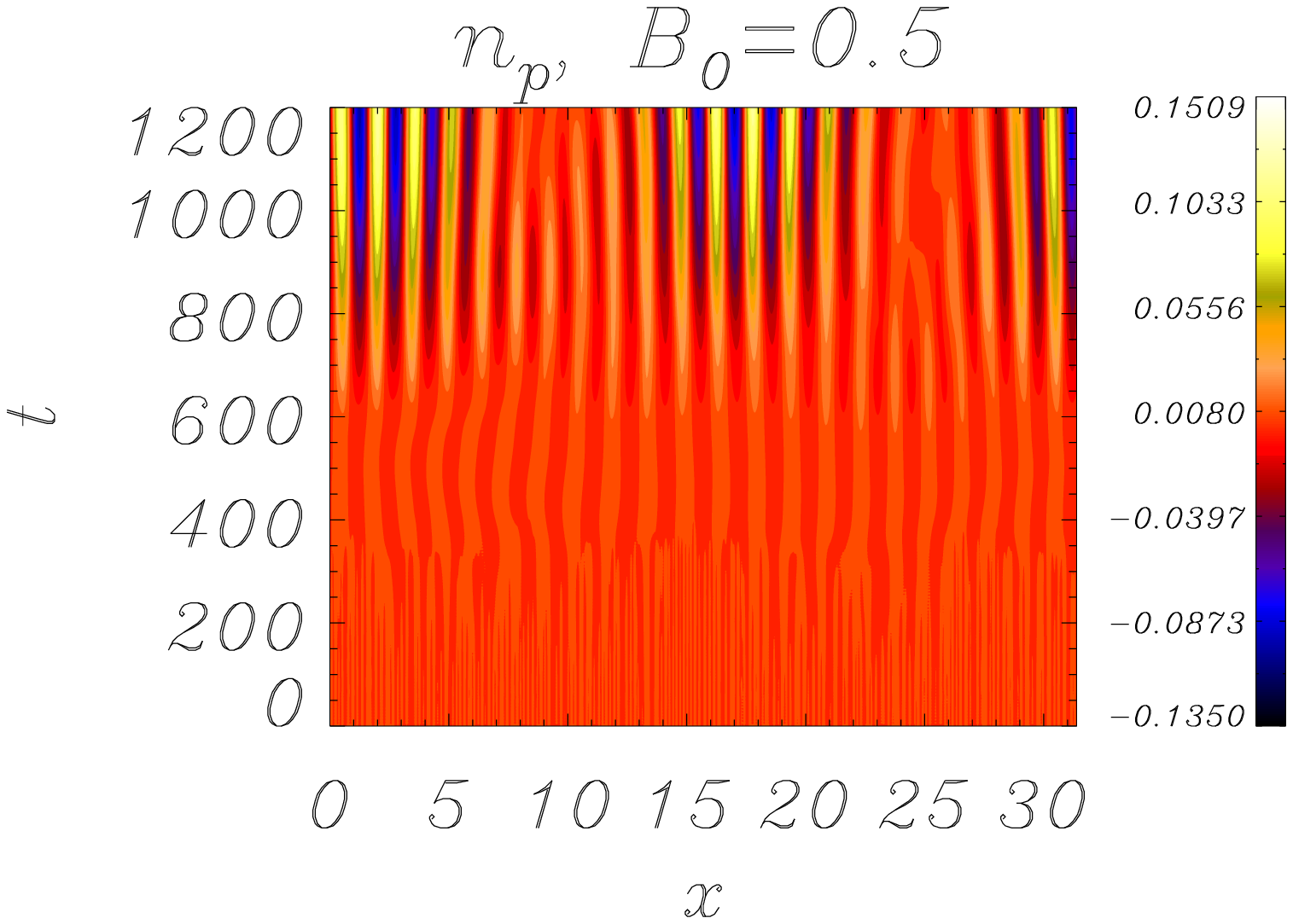,height=5.3cm,width=8.6cm}}
\vspace{-0.3cm}
\vspace{-0.1cm}\caption{\small Electron density and  proton density for $B_0=0.5$.}
\label{Fig10}
\end{figure}

\begin{figure}[!h]
\vspace{-0.0cm}
\centerline{\hspace{-0.2cm}\psfig{figure=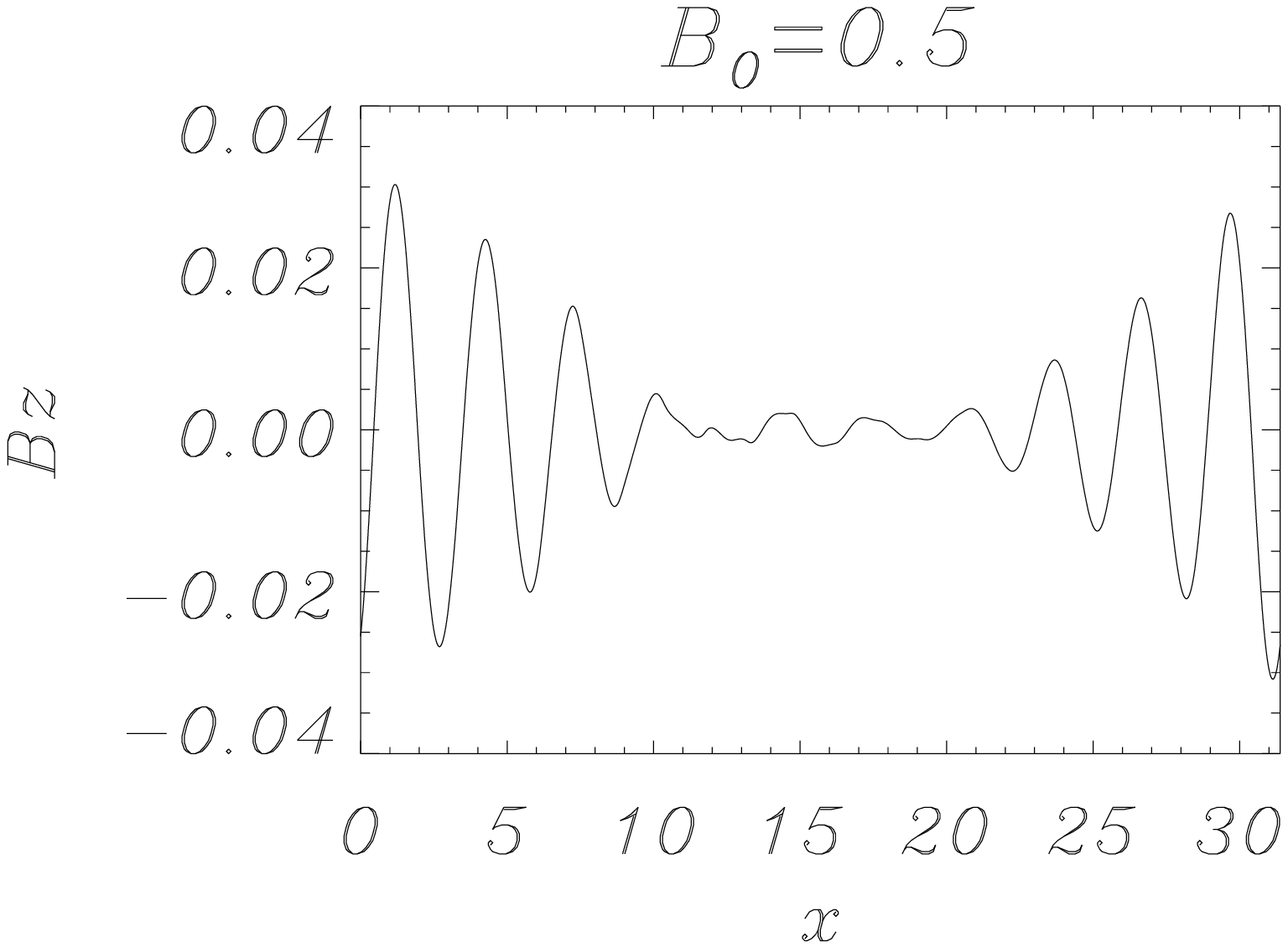,height=5.3cm,width=7.cm}
\hspace{-0.7cm}\psfig{figure=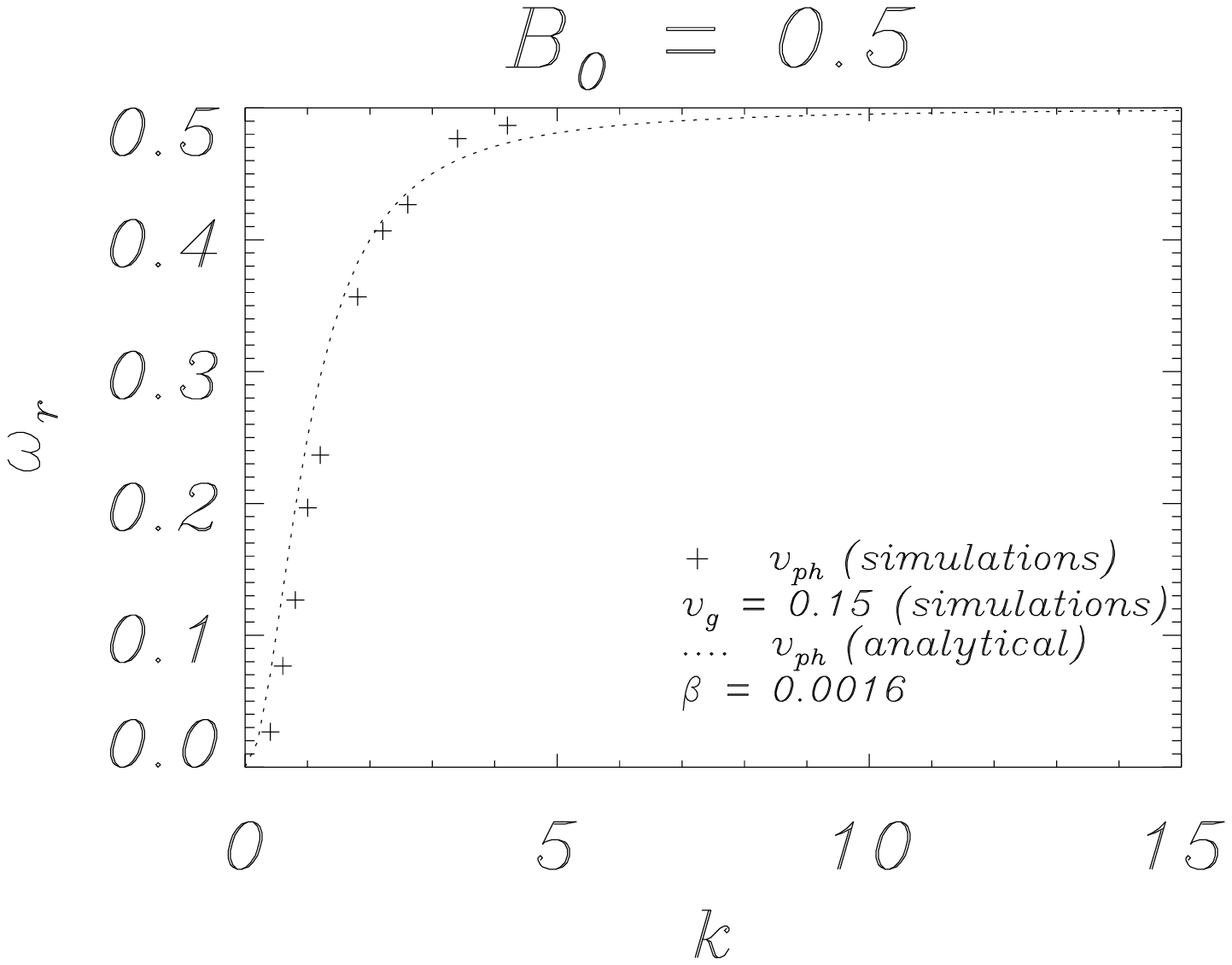,height=5.3cm,width=7.cm}}
\vspace{-0.4cm}\caption{\small Left frame: Magnetic field versus position at $t=540$. Right frame: The dispersion relation of  the whistler waves  available which have a value  of their phase velocity  close to that  of the corresponding  group velocity.}
\label{Fig11}
\end{figure}

{As we have already discussed in Sec.\ref{EF},  the distorted distribution function  for $B_0=0.1$  leads to the resonant excitation of short wavelength plasma waves  at  large wave numbers, e.g. ($k \sim 10$). On the contrary, at smaller wave numbers we see the  interplay  between the  inhomogeneity of the perturbed magnetic field amplitude  and of  the  electron and proton densities.} The electrostatic and the  magnetic fields are shown in $x$-$t$ space in Fig.\ref{Fig9} for $B_0=0.5$.  These figures show the forward and the backward propagation of the  whistler waves. In addition, long  wavelength  modulations  of the electrostatic structures and of the proton and electron densities are seen in Fig.\ref{Fig10}  at  half the wavelength  of the perturbed magnetic field.
 The wavelength of the modulations of both the magnetic field and  of the proton and electron densities  decreases with increasing  ambient magnetic field, the electrostatic field and density modulation wavelength  still  being equal to  half the wavelength of the most unstable Fourier component of the perturbed  magnetic field. We focus our attention  primarily  on  the case with ambient magnetic field $B_0 = 0.5$.
\\
{There are two kinds of modulations in the perturbed magnetic field, one with  long wavelengths having a width almost half the total length of the simulation box, $L \sim 17$ and corresponding time period $T \sim 200$. The phase velocity of these modulations is  $v_{ph} \sim 0.08$. The other kind has  shorter wavelengths,  $L \sim 2$, and  $T \sim 7$, i.e. $v_{ph} \sim 0.28$. The range of phase velocities  corresponding to  the peak in the whistler waves frequency spectrum for different values of $k$  extends from $0.005 \le v_{ph}$ to   $ v_{ph}\le 0.25$. Hence, the long wavelength phase velocity $v_{ph} \sim 0.08$, as obtained from the magnetic field modulations, falls in the range of the whistler phase velocities and it can be  further noticed that it is  comparable with some of  the phase velocities obtained for both short wavelength and long wavelength electrostatic modulations which are in the range $0.005 \le v_{ph} \le 0.01$ and  $0.075 \le v_{ph} \le 0.13$, respectively. Furthermore the range of phase velocities $0.08 \le v_{ph} \le 0.01$   obtained from low frequency peak, $0.03 \le \omega_r \le 0.04$ in the frequency spectrum of the electrostatic field at the most unstable mode $k_{max} = 3.6$ (see Fig.\ref{Fig2} bottom right frame) covers a large part of  the  phase velocity range of   the whistler waves. }
{We can thus infer that the dominant whistler waves are  modulated by the low frequency perturbations observed in the electrostatic field. This  kind of modulations has been studied theoretically.  In Ref.\onlinecite{Tripathy} it was proposed that  the nonlinear coupling of oppositely travelling waves (forward and backward waves) produce  a low frequency ponderomotive force with frequency $\omega = \omega_1 - \omega_2$ and wave vector $k = k_1 - k_2 = 2k_1$.  In Refs. \onlinecite{Eliasson} and \onlinecite{Stenflo}  it was later shown  that these ponderomotive forces are reinforced due to the interaction of the whistler packets with the background low density perturbations ~\cite{Eliasson}. In addition to the low density perturbations observed in our simulations (see Fig.\ref{Fig10}),  the formation of wave packets of the perturbed magnetic field  are also evident (see Fig.\ref{Fig11}, left frame). Coherent wave emission in the whistler frequency range $(0.1-0.5\omega_{ce})$, consisting of nearly monochromatic wave packets, has recently been observed by several spacecraft in the Earth's plasma environment \cite{Dubinin}. These are mainly generated by  oscillitons which are  stationary, nonlinear structures exhibiting solitary structures with embedded smaller-scale oscillations resembling wave packets. These arise from the momentum exchange between protons and electrons. These oscillitons in the whistler frequency range are known as {\em 'whistler oscillitons'}. The appearance of stationary nonlinear waves is related to the existence of a 'resonance point' where phase and group velocity coincide. The right frame in Fig.\ref{Fig11} shows the range of  frequencies and wave number available from the numerical results which have a value of their phase velocity  close to that  of the corresponding  group velocity.  We also observe a shift of the whistler waves to longer wavelength corresponding to the shift of the broad magnetic spectrum to lower $k$ values during the saturation regime when the coherent whistler emission takes place and modifies the frequency range of the whistler modes. \\ Observations from laboratory experiments \cite{Kostrov} exhibit a clear evidence of this kind of modulated whistler wave packets due to nonlinear effects. Furthermore, instruments on board CLUSTER spacecraft have observed \cite{Moullard}  broadband intense electromagnetic waves. Cluster measurements also exhibit the formation of envelop solitary waves accompanied by plasma density cavities. Both electron and proton density cavities  formed in our numerical results are shown in Fig.\ref{Fig10}.}

\section{Conclusions}

The anisotropy of  the electron distribution function ( $T_{\perp}>T_{||}$) has been shown to lead  to the onset of both resonant,  $\omega_r \geq \gamma_{max}$,  and non resonant,   $\omega_r << \gamma_{max}$, instabilities depending on the value of the ambient magnetic field, see Table  \ref{table1}.
When the plasma is unmagnetized the instability (Weibel instability) is completely nonresonant, where $\omega_r=0 << \gamma_{max}=0.038$ (in units of the Langmuir frequency), but as we move towards higher ambient magnetic fields,  $\omega_r$ becomes comparable to the maximum growth rate of the instability. At very high ambient magnetic field ($B_0=0.5$, being the value of the electron cyclotron frequency normalized to the Langmuir frequency) the mode becomes resonant with $\omega_r$ much larger than the growth rate (whistler instability).   \\The development of these instabilities is accompanied by the generation of electrostatic and electromagnetic waves at the  harmonics of the plasma frequency (up to the fifth harmonic in the case of the electrostatic spectrum). \\ In addition the spatial self-organization of the perturbed magnetic field leads to the deformation of the electron distribution function and to the consequent generation of short wavelength electrostatic modes. These secondary  phenomena tend to be suppressed  as the ambient magnetic field becomes larger.
On the contrary, at large ambient magnetic field we find a low frequency long wavelength modulation of the whistler wave spectrum that we interpret in terms of the self-consistent interaction between the perturbed magnetic field and the low frequency electron and proton density modulations.  The magnetic modulations that we observe are reminiscent of the so called ``whistler oscillitons''  that arise  under the condition that the mode phase velocity coincides with its group velocity.

The simulation results presented here provide relevant information on the different processes that the presence of anisotropy can drive in collisionless  plasmas in different plasma magnetization regimes, as parametrized by the value of the parameter $\beta_{\perp} > 1$ that gives the ratio between the perpendicular plasma pressure and the ambient magnetic field pressure. For example, in the context of solar physics,  the occurrence of  resonant whistler modes  has been advocated as an effective acceleration mechanisms  of  electrons directly from the background plasma up to $\sim 25 $ MeV within several seconds under  normal solar  flare conditions \cite{Miller}.  

{The simulation results presented in this paper have been obtained  by using a 1D configuration and are  thus restricted  to the  study of  the nonlinear effects on the system only of modes  with parallel wavenumber ${\bf k}$.  This restriction will be lifted in a future paper.
However, linear analysis 
\cite{Gary2000}
indicates that   in the presence of low ambient magnetic fields the whistler 
instability remains  the fastest growing one even in a two dimensional configuration where we allow modes to propagate at an angle with respect to the ambient field.  On the contrary, at higher ambient magnetic fields where the  whistler  instability  is suppressed,  obliquely propagating modes have a larger growth rate.  An important question to address in this  latter case  is how  the formation of  electrostatic structures will develop in  such a regime.
}

\section*{Acknowledgements} 
{ The numerical calculations presented in this work were performed at the Italian Super-computing Center {\em Cineca} (Bologna), supported by Cineca and by CNR-INFM. }

\end{document}